\documentclass[
  aps,
  prb,
  twocolumn,
  superscriptaddress,
  amsmath,amssymb,
  floatfix,
  nofootinbib
]{revtex4-2}

\usepackage{graphicx}
\usepackage{bm}
\usepackage{braket}
\usepackage{amsmath}
\usepackage{mathtools}
\usepackage{xcolor}
\usepackage{booktabs}
\usepackage[colorlinks=true,linkcolor=blue,citecolor=blue,urlcolor=blue]{hyperref}

\newcommand{\Sv}{\bm{S}}
\newcommand{\nv}{\hat{\bm{n}}}
\newcommand{\Dv}{\bm{D}}
\newcommand{\Bv}{\bm{B}}
\newcommand{\rv}{\bm{r}}
\newcommand{\Rmat}{\mathcal{R}}
\newcommand{\Heff}{H_{\mathrm{eff}}}
\newcommand{\Ush}{U_{\mathrm{sh}}}
\newcommand{\Fc}{F_{c}}
\newcommand{\SUtwo}{\mathrm{SU}(2)}
\newcommand{\SOthree}{\mathrm{SO}(3)}
\newcommand{\Trace}{\mathrm{Tr}}
\newcommand{\Tr}{\mathrm{Tr}}
\newcommand{\rpar}{r_{\parallel}}
\newcommand{\ELZ}{E_{\mathrm{LZ}}}
\DeclareMathOperator{\artanh}{artanh}
\newcommand{\Sisotope}{$^{28}$Si}
\newcommand{\Jiso}{J_{\mathrm{iso}}}
\newcommand{\eq}[1]{Eq.~(\ref{#1})}
\newcommand{\fig}[1]{Fig.~\ref{#1}}
\newcommand{\sect}[1]{Sec.~\ref{#1}}

\begin{document}

\title{Spin Shuttling Through Quantum-Dot Based Chiral Molecular Bridges: Gauge Constraints, Holonomy, and Quantum Gates}

\author{Amitoj S. Miglani}
\thanks{These authors contributed equally to this work.}
\affiliation{Department of Physics, Indian Institute of Technology Roorkee, Roorkee 247667, India.}

\author{Rajdeep R. Dwivedi}
\thanks{These authors contributed equally to this work.}
\affiliation{Department of Electronics and Communication Engineering, Indian Institute of Technology Roorkee, Roorkee 247667, India.}

\author{Balsant S. Tiwari}
\affiliation{University of Birmingham, Birmingham, B15 2TT, United Kingdom.}

\author{Vishvendra S. Poonia}
\email{vishvendra@ece.iitr.ac.in}
\affiliation{Department of Electronics and Communication Engineering, Indian Institute of Technology Roorkee, Roorkee 247667, India.}

\date{\today}

\begin{abstract}
Chiral molecular bridges between silicon quantum dots have been proposed to furnish novel two-qubit entangling resources via the chirality-induced spin-selectivity (CISS) effect. We evaluate this premise using Makhlin invariants. We model a phase-coherent bridge as an $\mathrm{SU}(2)$-valued hopping between two spins and apply a Schrieffer--Wolff reduction which yields a twisted exchange interaction $J\,\Sv_1\!\cdot\!\Rmat\,\Sv_2$, whose Dzyaloshinskii--Moriya and symmetric anisotropic parts are both fixed by a single bridge angle $\alpha$. Using the Kaplan--Shekhtman--Entin-Wohlman--Aharony (KSEA) gauge argument, we prove this interaction is locally equivalent to isotropic exchange, showing that a single chiral bond does not constitute a new entangling class. We identify genuine chiral utility that emerges once this gauge symmetry is broken. Chiral exchange converts a uniform transverse field into a synthetic gradient, bringing the two-qubit propagator to within $7\times10^{-4}$ of the CNOT class in invariant space and achieving this 4.7 times faster than an explicit orthogonal gradient of equal magnitude. Shuttling a spin through bridges with non-collinear axes implements a non-Abelian $\SUtwo$ holonomy, yielding rotationally non-reciprocal transport and closed-loop geometric gates that suppress dynamical-phase noise to second order. The microscopic origin of CISS is still unsettled. To address this, we introduce a unified transport channel combining coherent rotation, a direction-reversing spin filter, and pointer-basis dephasing. This model obeys an exact $\tanh$ composition law for pumped polarisation and a closed-form transverse decay law, yielding four complementary observables that experimentally over-determine the bridge's coherent fraction $\Fc$. Finally, we quantify silicon-specific hardware constraints, establishing the required Landau--Zener valley-adiabaticity window ($E_v\gtrsim2.1\sqrt{\hbar v}$) and charge-noise limit ($\sigma_J/J\lesssim3\%$) for $99\%$ gate fidelity.
\end{abstract}

\maketitle

\section{Introduction}
\label{sec:intro}
Electron spins confined in gate-defined silicon quantum dots are a leading
platform for scalable quantum information processing. Isotopically purified
\Sisotope\ hosts spin qubits with second-scale relaxation times and demonstrated
single- and two-qubit fidelities above the surface-code
threshold~\cite{Zwanenburg2013,Veldhorst2014,Yoneda2018,Noiri2022,Xue2022,Burkard2023,Broz2026AEON,HRL2026QPU}.
A central architectural challenge is connectivity, namely moving quantum
information coherently between distant dots without a continuous chain of
nearest-neighbour exchange gates. Coherent spin shuttling (physically
transporting an electron and its spin through an array) has emerged as the leading approach, with demonstrations of charge and spin transfer through
linear dot arrays and conveyor-mode
channels~\cite{Fujita2017,Mills2019,Yoneda2021,Seidler2022,Zwerver2023,Langrock2023}.

In parallel, the chirality-induced spin-selectivity (CISS) effect, the
observation that electron transmission through chiral
molecules is spin dependent even in the absence of magnetic
elements~\cite{Gohler2011,Naaman2012,Naaman2019,Evers2022}, has raised the
prospect of a purely molecular route to spin control. If a chiral
molecular bridge linking two silicon dots transmits the two spin projections
differently, interdot tunnelling becomes spin dependent, and one might hope to
engineer novel exchange couplings, non-reciprocal spin transport, and new
entangling gates. This expectation motivates work at the interface of molecular spintronics and quantum information~\cite{Aiello2022}, but it has not been tested against the symmetry constraints on two-qubit gates.

In this work, we evaluate these arguments against the symmetry constraints governing two-qubit gates and explore the resulting architectural implications. We first classify the single-bond case. A single chiral
bond, treated as a coherent spin-dependent ($\SUtwo$) hopping, generates
no new two-qubit entangling class. The effective spin Hamiltonian is a twisted (Dzyaloshinskii--Moriya-carrying) exchange interaction
that is locally unitarily equivalent to ordinary isotropic Heisenberg
exchange, representing the two-site instance of the
Kaplan--Shekhtman--Entin-Wohlman--Aharony (KSEA) gauge theorem of anisotropic
superexchange~\cite{Dzyaloshinsky1958,Moriya1960,Kaplan1983,Shekhtman1992,Kavokin2001}.
While the KSEA theorem is well-established, we translate its implications into the operational framework of quantum information processing. We show at the
level of the Makhlin local invariants and the exact entangling
power that the chiral-bond and isotropic propagators are indistinguishable
at all times. Furthermore, we identify precisely which truncation produces the appearance of a continuously tunable gate family, specifically the dropping of the
symmetric anisotropy $\Gamma$ that is locked to the Dzyaloshinskii--Moriya
term.

Having established the operational limitations of a single chiral bond, we now delineate the specific regimes wherein chirality affords genuine computational utility. Chirality manifests as a distinct resource when the single-bond gauge freedom is broken by an additional non-commuting structure.
\begin{enumerate}
\item[(i)] \emph{Gauge criteria and the generation of a synthetic field gradient.}

Upon the introduction of local Zeeman fields, the two-qubit equivalence class depends on the bridge angle exclusively through the relative orientation between $\Bv_1$ and the back-rotated field $\Rmat_\alpha^{-1}\Bv_2$ (\sect{sec:zeeman}). This explicit criterion identifies geometric configurations wherein chirality remains mathematically inert, such as $\Bv_2\parallel\nv$, alongside configurations where it provides a substantial operational advantage. Specifically, under a uniform transverse field, chiral exchange generates a synthetic field gradient characterized by an angle exactly equal to $\alpha$. This mechanism drives the two-qubit propagator to within $7\times10^{-4}$ of the CNOT class in invariant space, achieving this proximity $4.7$ times faster than an explicit orthogonal gradient of equal field magnitude.
\item[(ii)] \emph{Non-Abelian shuttling holonomy.}

Transporting a spin through a sequence of bridges possessing mutually non-collinear chiral axes applies a path-ordered product of $\SUtwo$ rotations. The net geometric operation constitutes a genuine non-Abelian holonomy. This process is inherently direction-dependent, producing distinct forward and backward transport states due to rotational non-reciprocity. Furthermore, a closed shuttling loop implements a gauge-invariant Wilson-loop single-qubit gate. In accordance with the principles of holonomic quantum computation, this geometric gate effectively suppresses dynamical-phase noise to second order~\cite{Zanardi1999,Duan2001,Wilczek1984,Sjoqvist2015}.
\end{enumerate}

Spin rotation during transport is not unique to molecular bridges. Intrinsic
spin--orbit coupling rotates hole spins shuttled through germanium quantum
dots~\cite{VanRiggelen2024} and contributes coherently to conveyor-mode
transport in Si/SiGe~\cite{Struck2024,Langrock2023}. For conduction
\emph{electrons} in silicon, however, the intrinsic rotation per interdot
transfer is minute. An engineered bridge therefore remains the most realistic route to achieve $\mathcal{O}(1)$ rotation angles per hop. Relative to intrinsic spin--orbit coupling, a chiral
molecular bridge offers a rotation axis set by chemistry rather than by
strain and electric fields, sign-programmable by enantiomer selection, and
requiring no additional gate electrodes. The primary operational trade-off is the open question of
bridge coherence, which \sect{sec:channel} converts into a targeted measurement
programme.

Because the microscopic status of CISS remains
debated~\cite{Evers2022,Fransson2020,Dalum2019,Alwan2021}, we treat the two
dominant interpretations on an equal footing. First, we consider a \emph{coherent} $\SUtwo$
spin rotation imprinted on the tunnelling amplitude. Second, we consider an
\emph{incoherent} channel in which the bridge acts as a direction-reversing
spin filter and dephases the spin in a bridge-defined pointer basis. We
interpolate between them with a phenomenological \emph{coherent fraction}
$\Fc\in[0,1]$ and show that the two limits are experimentally distinguishable
through four complementary observables, including an exact,
direction-reversing polarisation-pumping law. We then specialise to silicon,
incorporating the valley degree of freedom, which has no analogue in the
molecular-spintronics literature but is decisive for any real \Sisotope\
device~\cite{Friesen2007,Culcer2010}, and provide a comprehensive charge-noise error
budget.

The physical assumptions underlying these results and their quantitative impacts are summarized in \sect{sec:limitations}. Our aim is not to claim a fully realized working chiral gate, but to delineate precisely which chiral effects survive symmetry analysis and how to rigorously measure them experimentally. To this end, the remainder of the paper formalizes the chiral-bridge model (\sect{sec:model}), derives the twisted-exchange Hamiltonian and gauge triviality (\sect{sec:gauge}), establishes the gauge criterion and holonomy resources (\sect{sec:zeeman}, \sect{sec:holonomy}), and constructs a measurement programme for the bridge coherence (\sect{sec:channel}) before concluding with an error budget accounting for silicon valley physics (\sect{sec:valley}, \sect{sec:budget}).


\section{Model of a chiral bridge as spin-dependent hopping}
\label{sec:model}

\subsection{Device and Hamiltonian}
\begin{figure*}[t]
\centering
\includegraphics[width=0.92\textwidth]{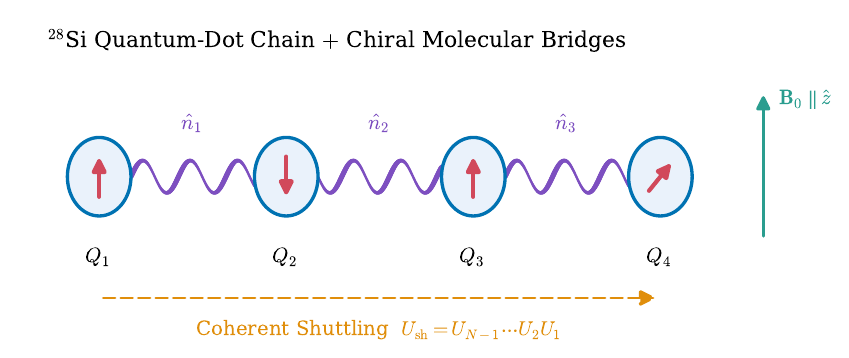}
\caption{Silicon quantum-dot chain connected by chiral molecular bridges.
Each dot $Q_i$ hosts a single electron spin, and a global field
$\Bv_0\parallel\hat z$ sets the quantisation and readout axis. Bridge $i$
imprints a spin rotation about its own chiral axis $\nv_i$ (helicity encoded
in the drawn handedness). Coherent shuttling applies the path-ordered product
$\Ush=U_{N-1}\cdots U_2U_1$. The holonomy is non-Abelian as the axes $\nv_i$ need not be collinear.}
\label{fig:schematic}
\end{figure*}

We consider a linear chain of gate-defined quantum dots in isotopically
enriched \Sisotope, where each dot hosts a single electron in its orbital/valley
ground state and is connected by chiral molecular linkers. For concreteness, we assume
helicene-type molecules chemically anchored to a hydrogen-terminated silicon
surface, as illustrated in \fig{fig:schematic}. A global magnetic field
$\Bv_0\parallel\hat{z}$ defines the qubit quantisation axis and the readout
basis. The relevant low-energy degrees of freedom are the localised spins
$\Sv_i$. The orbital charge is either static for exchange gates or
transported for shuttling operations.

The single-particle bridge hopping is written as a spin-dependent tunnelling
operator between neighbouring dots $i$ and $j$,
\begin{equation}
H_{\mathrm{hop}} = -t\,\big(c^{\dagger}_{i}\,U_{ij}\,c_{j} + \text{h.c.}\big),
\qquad U_{ij}\in\SUtwo,
\label{eq:hop}
\end{equation}
where $c_i=(c_{i\uparrow},c_{i\downarrow})^{T}$ and $t>0$ is the spin-independent
tunnelling amplitude. The operator $U_{ij}$ encodes the spin transformation imparted by
the chiral bridge. This represents the standard method by which spin-orbit-active bonds enter a
single-band Hubbard description~\cite{Kaplan1983,Kavokin2001,Baruffa2010}. We
parametrise this transformation as
\begin{equation}
U_{ij} = \exp\!\left(\tfrac{i}{2}\,\alpha_{ij}\,\nv_{ij}\!\cdot\!\bm{\sigma}\right)
= \cos\tfrac{\alpha_{ij}}{2}\,\mathbb{1}
+ i\sin\tfrac{\alpha_{ij}}{2}\,\nv_{ij}\!\cdot\!\bm{\sigma},
\label{eq:Uij}
\end{equation}
with $\bm{\sigma}=(\sigma_x,\sigma_y,\sigma_z)$ representing the Pauli vector and $\nv_{ij}$ denoting
the unit chiral axis. This axis is fixed by the molecular geometry and reverses sign with the handedness of the molecule. The parameter $\alpha_{ij}$ is the
$\SOthree$ rotation angle acquired in transit. The corresponding
Bloch-sphere ($\SOthree$) rotation is
$\Rmat_{ij}=\exp(\alpha_{ij}\,\nv_{ij}\cdot\bm{L})$ with $(\bm{L}_a)_{bc}
=-\epsilon_{abc}$. The factor of two between the $\SUtwo$ half-angle in
\eq{eq:Uij} and the physical $\SOthree$ rotation angle is retained explicitly to distinguish operations in the two spaces.

\subsection{Two interpretations of CISS}
The microscopic origin of CISS is not settled, and different mechanisms imply
different mathematical structures for the bridge action. We therefore evaluate two
distinct scenarios in parallel.

In the first scenario, the bridge acts as a purely coherent unitary spin rotation defined by \eq{eq:Uij} with a real $\alpha_{ij}$ and a definite axis. Physically, this
corresponds to spin--orbit precession during a phase-coherent traversal, similarly to
a helical Rashba channel~\cite{Yeganeh2009,Guo2012}. A unitary rotation is trace
preserving with a spin-independent transmission probability, meaning it imparts
no net polarisation to an unpolarised electron. Strictly coherent CISS therefore
produces none of the transmission polarisation that defines the effect
experimentally. The large room-temperature polarisations reported in the literature imply a substantial nonunitary component in those specific regimes. Whether the
cryogenic, single-electron, phase-coherent regime relevant to quantum dot arrays behaves differently
is precisely the question the protocols of \sect{sec:channel} are designed to decide.

In the second scenario, the bridge acts as an incoherent spin filter. It
transmits the two eigenstates of $\nv_{ij}\!\cdot\!\bm{\sigma}$ with
different probabilities and polarises the transmitted spin along $\pm\nv_{ij}$
with the sign set by the propagation direction due to helicity-momentum locking. When which-path information is recorded by the environment, the bridge
additionally dephases the spin in the $\nv_{ij}$ pointer basis. The reduced
dynamics, conditioned on successful transfer, is then a nonunitary map
rather than a rotation. This matches the phenomenology of standard room-temperature
CISS transport~\cite{Gohler2011,Naaman2019,Fransson2019}.

\subsection{Unified description and the coherent fraction $\Fc$}
\label{sec:unified}
Real bridges exist between these two extreme limits. The physically correct
interpolation must contain the spin filter of the second scenario rather than merely simple dephasing. We model the one-bridge channel for propagation direction
$s=\pm1$ as a filter, followed by pointer-basis dephasing, and then by the
coherent rotation. This sequence reflects a physical model where spin-dependent transmission precedes environmental dephasing and coherent precession,
\begin{equation}
\mathcal{E}^{(s)}_{ij}(\rho) = U_{ij}^{\,s}\,
\Lambda_{q}\!\Big[\frac{M_{s}\,\rho\,M_{s}}{\Tr(M_{s}\rho M_{s})}\Big]
\,U_{ij}^{s\dagger},
\label{eq:channel}
\end{equation}
utilising the unnormalised Kraus filter and the dephasing map
\begin{align}
M_{s}&=\sqrt{\tfrac{1+s\kappa}{2}}\,P_{+}
      +\sqrt{\tfrac{1-s\kappa}{2}}\,P_{-},
\label{eq:filter}\\
\Lambda_{q}(\rho)&=(1-q)\rho+q\,\big(P_{+}\rho P_{+}+P_{-}\rho P_{-}\big),
\end{align}
where $P_{\pm}=\tfrac12(\mathbb{1}\pm\nv_{ij}\!\cdot\!\bm{\sigma})$ project on
$\nv_{ij}\!\cdot\!\bm{\sigma}=\pm1$. The parameters are defined as
\begin{equation}
\kappa=(1-\Fc)\,\lambda_{f},\qquad q=(1-\Fc)\,\lambda_{d}.
\label{eq:kq}
\end{equation}
Here $\kappa\in[0,1]$ is the filtering polarisation and $q\in[0,1]$ is the
dephasing depth, while $\lambda_{f}$ and $\lambda_{d}$ set their fully incoherent
($\Fc=0$) maxima. The rotational operator $U^{s}_{ij}=\exp(-\tfrac{i}{2}s\alpha_{ij}\,
\nv_{ij}\!\cdot\!\bm{\sigma})$ reverses with direction. The sign of the
filter tracks $s$, meaning forward and backward traversals prefer opposite spins. At $\Fc=1$ the map is a pure rotation, whereas at $\Fc=0$ it becomes a maximally filtering and maximally dephasing bridge.

Equation~(\ref{eq:channel}) is exactly solvable on the Bloch sphere. Writing
$\rv=\rpar\nv+\rv_{\perp}$ with $\rpar=\nv\!\cdot\!\rv$, one traversal gives
\begin{align}
\rpar &\mapsto \frac{s\kappa+\rpar}{1+s\kappa\,\rpar},
\label{eq:mobius}\\
\rv_{\perp} &\mapsto \frac{(1-q)\sqrt{1-\kappa^{2}}}{1+s\kappa\,\rpar}\;
\mathcal{R}_{\nv}(s\alpha)\,\rv_{\perp}.
\label{eq:transverse}
\end{align}
This represents a M\"obius map on the longitudinal component and a contraction-plus-rotation
on the transverse component. In rapidity variables $\theta=\artanh\rpar$ and
$a=\artanh\kappa$, \eq{eq:mobius} linearises to
$\theta\mapsto\theta+s\,a$. After $N$ identical bridges, the pumped
polarisation follows an exact composition law,
\begin{equation}
\rpar^{(N)}=\tanh\!\big(\artanh\rpar^{(0)}+sN a\big),
\label{eq:tanh}
\end{equation}
and telescoping the prefactor in \eq{eq:transverse} yields
\begin{equation}
\big|\rv_{\perp}^{(N)}\big|=\big|\rv_{\perp}^{(0)}\big|\,(1-q)^{N}\,
\frac{\cosh\theta_{0}}{\cosh(\theta_{0}+sN a)},
\label{eq:perpN}
\end{equation}
with $\theta_{0}=\artanh\rpar^{(0)}$.

The filter acts as a polariser of rapidity $a$ per bridge. Because its fixed
points are the pure states $\rv=\pm\nv$, the state purity is
non-monotonic. This demonstrates that state purity alone is an insufficient witness for coherent transport (\sect{sec:channel}). A single traversal of an
unpolarised input yields $\rpar^{(1)}=s\kappa$, demonstrating that $\kappa$ is exactly the
transmission polarisation measured in standard CISS experiments. The variables $(\Fc,\lambda_{f},\lambda_{d})$ are phenomenological. Existing CISS data
constrain only the product $(1-\Fc)\lambda_{f}=\kappa$, and deriving these
parameters from a microscopic transport calculation is left for future work
(\sect{sec:limitations}). The complete derivation is provided in Appendix~\ref{app:channel}.

\section{Twisted exchange and the gauge triviality of a single bond}
\label{sec:gauge}

\subsection{Effective two-spin Hamiltonian}
\begin{figure*}[t]
\centering
\includegraphics[width=0.92\textwidth]{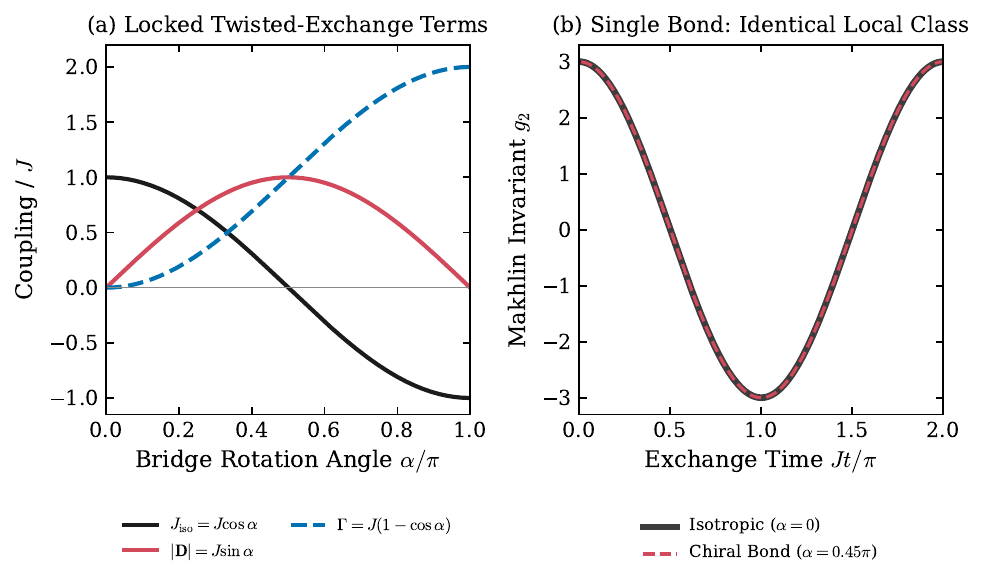}
\caption{Gauge triviality of a single chiral bond. (a) The three couplings in
the twisted-exchange Hamiltonian (\ref{eq:decomp}) are locked to the single
bridge angle $\alpha$. These include the isotropic exchange $\Jiso=J\cos\alpha$, the DM magnitude
$|\Dv|=J\sin\alpha$, and the symmetric anisotropy $\Gamma=J(1-\cos\alpha)$. (b) The Makhlin invariant $g_2$ of the
propagator $\exp(-i\Heff t)$ is plotted as a function of exchange time for a chiral bond
($\alpha=0.9\times\pi/2$, dashed red) and for isotropic exchange
($\alpha=0$, solid black). The curves coincide to machine precision
[$\max|g_2^{\mathrm{chiral}}-g_2^{\mathrm{iso}}|=4\times10^{-15}$], proving
that a single chiral bond generates the same set of local-equivalence
classes as isotropic exchange.}
\label{fig:gauge}
\end{figure*}

In the exchange (static-charge) regime, two electrons in neighbouring dots at
half filling interact through virtual tunnelling across the bridge. A
second-order Schrieffer-Wolff transformation~\cite{Bravyi2011} of the
extended Hubbard model with the spin-dependent hopping (\ref{eq:hop}) yields
the effective Hamiltonian derived in Appendix~\ref{app:sw},
\begin{equation}
\Heff = J\,\Sv_1\!\cdot\!\big(\Rmat\,\Sv_2\big),
\qquad J=\frac{4t^{2}}{U_H},
\label{eq:twisted}
\end{equation}
where $U_H$ is the on-site charging energy, $\Rmat=\Rmat_{12}\in\SOthree$ is
the rotation associated with $U_{12}$ through
$U\,(\bm{a}\!\cdot\!\bm\sigma)\,U^{\dagger}=(\Rmat\bm{a})\!\cdot\!\bm\sigma$,
and $\Sv=\tfrac12\bm\sigma$. Equation (\ref{eq:twisted}) represents the
\emph{twisted} (or Moriya-Kaplan) exchange interaction. Using the Rodrigues
decomposition of $\Rmat$ about axis $\nv$ by angle $\alpha$, we obtain
\begin{align}
\Heff &= \underbrace{J\cos\alpha}_{\Jiso}\,\Sv_1\!\cdot\!\Sv_2
\;+\; \underbrace{J\sin\alpha}_{|\Dv|}\;\nv\!\cdot\!(\Sv_2\times\Sv_1)
\nonumber\\
&\quad +\; \underbrace{J(1-\cos\alpha)}_{\Gamma}\,(\nv\!\cdot\!\Sv_1)(\nv\!\cdot\!\Sv_2),
\label{eq:decomp}
\end{align}
so that the antisymmetric Dzyaloshinskii-Moriya (DM) vector is
$\Dv = J\sin\alpha\,\nv$, the isotropic exchange is $\Jiso=J\cos\alpha$, and
there is an unavoidable \emph{symmetric-anisotropic} term of strength
$\Gamma=J(1-\cos\alpha)$. The three couplings are not independent. They are
locked to the single geometric angle $\alpha$ according to
\begin{equation}
\frac{|\Dv|}{\Jiso}=\tan\alpha,\qquad
\frac{\Gamma}{\Jiso}=\sec\alpha-1 ,
\label{eq:locked}
\end{equation}
as shown in \fig{fig:gauge}(a). This locking is the central mechanism of the gauge argument.
One cannot have the DM term without the symmetric anisotropy, and it is precisely this symmetric anisotropy that restores local equivalence to the Heisenberg exchange.
We explicitly define our conventions for this derivation. The antisymmetric part of \eqref{eq:decomp} is $+J\sin\alpha\,\boldsymbol{n}\cdot(\boldsymbol{S}_{2}\times\boldsymbol{S}_{1}) = -\boldsymbol{D}\cdot(\boldsymbol{S}_{1}\times\boldsymbol{S}_{2})$ with $\boldsymbol{D}=J\sin\alpha\,\boldsymbol{n}$. We use the $(\boldsymbol{S}_{1}\times\boldsymbol{S}_{2})$ ordering throughout.

\subsection{The gauge transformation}
Because $\Rmat\in\SOthree$, we may write $\Rmat=e^{\alpha\,\nv\cdot\bm L}$ and
absorb it via a \emph{local} single-spin rotation. We define the site-2
rotation $V_2=\exp(-\tfrac{i}{2}\alpha\,\nv\!\cdot\!\bm\sigma)$ acting exclusively on spin
$2$. The effective Hamiltonian then transforms as detailed in Appendix~\ref{app:gauge},
\begin{equation}
V_2^{\dagger}\,\Heff\,V_2 = J\,\Sv_1\!\cdot\!\Sv_2 ,
\label{eq:gaugeaway}
\end{equation}
where the twisted exchange is transformed exactly into isotropic
Heisenberg exchange by a single-qubit unitary. This represents the
Kaplan-Shekhtman-Entin-Wohlman-Aharony
result~\cite{Kaplan1983,Shekhtman1992} in the two-site setting. A bond DM
interaction that is not frustrated by competing bonds is a gauge artefact, which is entirely
removable by a local spin-frame redefinition. Since local unitaries are
precisely the free operations in the entangling-power~\cite{Zanardi2000} and Makhlin
classification of two-qubit gates~\cite{Makhlin2002,Zhang2003}, a propagator
generated by $\Heff$ lies in the \emph{same} local-equivalence class as one
generated by isotropic exchange at the corresponding time. Both generate only the
one-parameter family through SWAP and $\sqrt{\text{SWAP}}$.

\subsection{Numerical confirmation via Makhlin invariants}
The Makhlin invariants $(g_1,g_2)$ are functions of a two-qubit unitary that
are invariant under $U\to (u_1\otimes u_2)\,U\,(v_1\otimes v_2)$ with
$u_i,v_i\in\SUtwo$. They uniquely label local-equivalence classes, as outlined in
Appendix~\ref{app:makhlin}. We compute these invariants for
$U(t)=\exp(-i\Heff t)$. \fig{fig:gauge}(b) plots $g_2(t)$ for a chiral bond
with $\alpha=0.9\times\pi/2$ against the isotropic ($\alpha=0$) result. The
two curves are indistinguishable with a maximum numerical deviation of
\begin{equation}
\max_{t}\,\big|g_2^{\mathrm{chiral}}(t)-g_2^{\mathrm{iso}}(t)\big|
= 4.0\times10^{-15} .
\end{equation}
The equality is mathematically exact and does not need to be
argued statistically. The Haar-averaged entangling
power~\cite{Zanardi2000,Makhlin2002} of a two-qubit gate depends only on the
local invariant $g_1$ through the relation
\begin{equation}
e_{p}(U)=\tfrac{2}{9}\big(1-|g_1(U)|\big) .
\label{eq:ep}
\end{equation}
Because gauge equivalence preserves $g_1$, it forces $e_p$ to coincide for the
chiral and isotropic bonds at every exchange time, both equaling $1/6$ at the
$\sqrt{\mathrm{SWAP}}$ point. A paired Monte-Carlo evaluation over $4000$
shared random product states confirms this theoretical constraint, yielding $0.1707$ versus $0.1678$ with a
paired standard error of $0.0016$. This forms the quantitative statement of the
gauge argument, proving that a single chiral bond is not a new entangling resource.

\subsection{Origin of the apparent continuously tunable gate family}
We briefly outline how conflicting conclusions can arise from partial Hamiltonians. If one retains only the isotropic and DM parts of \eq{eq:decomp}
while dropping the symmetric anisotropy $\Gamma$, the resulting truncated Hamiltonian
$J\Sv_1\!\cdot\!\Sv_2+\Dv\!\cdot\!(\Sv_1\times\Sv_2)$ is \emph{not} local-unitary
equivalent to Heisenberg exchange. This truncation typically occurs when treating
$\Dv$ as an independent small parameter and working to first order in $\Dv/J$
while neglecting $O(\Dv^2/J)$ terms. Under these approximations, the Makhlin invariants of the system genuinely move
with the DM angle. The apparent continuously tunable gate family is
therefore a mathematical artefact of an inconsistent truncation. At the order at which
$\Dv$ appears, $\Gamma$ also appears, and it is exactly the coefficient that
cancels the nonlocality. A consistent single-bond treatment must retain all three
terms in (\ref{eq:decomp}). When all terms are included, the chiral bond cleanly collapses onto the
Heisenberg class.

\section{A gauge criterion for two-qubit classes and chirality as a synthetic
gradient}
\label{sec:zeeman}
The single-bond gauge freedom is broken as soon as a second non-commuting
term is present that cannot be simultaneously rotated away. The most direct method to break this gauge symmetry is the introduction of a site-dependent Zeeman field,
\begin{equation}
H_Z=\Bv_1\!\cdot\!\Sv_1+\Bv_2\!\cdot\!\Sv_2 .
\end{equation}
Applying the same local rotation $V_2$ that removes the twist from $\Heff$
in \sect{sec:gauge} maps the system according to
\begin{equation}
\Heff+H_Z \;\xrightarrow{\;V_2\;}\;
J\,\Sv_1\!\cdot\!\Sv_2+\Bv_1\!\cdot\!\Sv_1
+\big(\Rmat_\alpha^{-1}\Bv_2\big)\!\cdot\!\Sv_2 .
\label{eq:gauged}
\end{equation}
This transformed Hamiltonian represents an ordinary Heisenberg-plus-gradient problem with a
\emph{back-rotated} site-2 field. All chiral effects are entirely absorbed into
that single geometric rotation. Consequently, the two-qubit local-equivalence class depends on the
bridge angle $\alpha$ exclusively through the relative field geometry
\begin{equation}
\theta_{\mathrm{eff}}=\angle\big(\Bv_1,\;\Rmat_\alpha^{-1}\Bv_2\big).
\label{eq:criterion}
\end{equation}
This criterion yields two immediate corollaries. First, if $\Bv_2\parallel\nv$, then
$\Rmat_\alpha^{-1}\Bv_2=\Bv_2$ and the class is rigorously
$\alpha$-independent. Under this condition, chirality exerts no operational effect regardless of the specific field magnitudes.
Second, if the field is uniform and transverse such that $\Bv_1=\Bv_2=b\,\hat e\perp
\nv$, then $\theta_{\mathrm{eff}}$ equals $\alpha$ exactly. Here, the chiral exchange converts a
homogeneous field into a synthetic gradient of a chemically programmable angle.
The first corollary explains why a configuration with $\Bv_2\parallel\nv$
cannot exhibit a chiral effect, while the second corollary transforms the underlying gauge structure
into a computational resource.

\begin{figure*}[t]
\centering
\includegraphics[width=\textwidth]{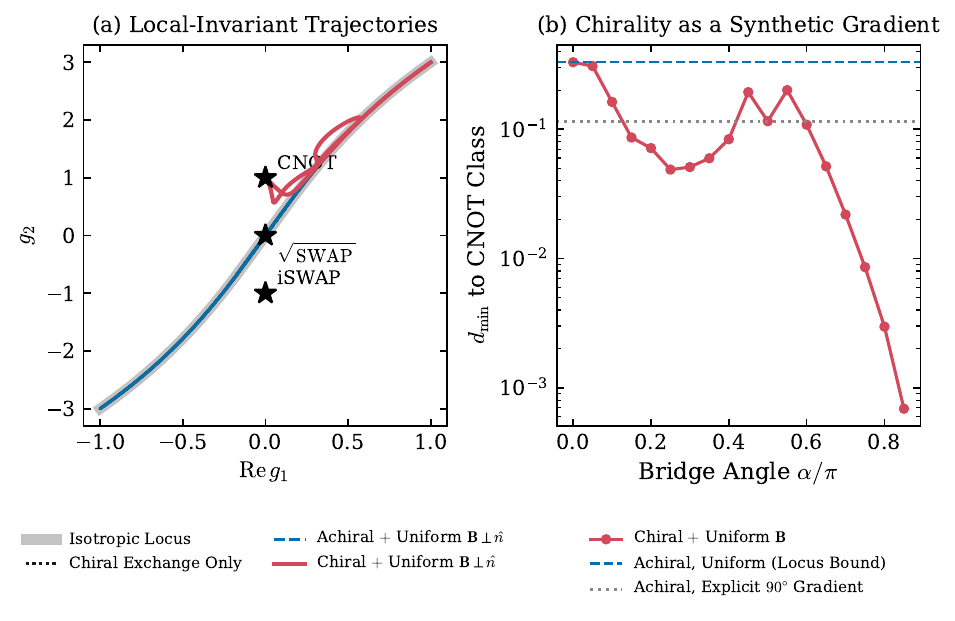}
\caption{Chirality as a synthetic field gradient. (a) Local-invariant
trajectories are plotted in the $(\mathrm{Re}\,g_1,g_2)$ plane for
$U(t)=\exp[-i(\Heff+H_Z)t]$ with a \emph{uniform} transverse field
$\Bv_1=\Bv_2=b\,\hat x\perp\nv$ ($\nv=\hat z$, $b=0.9J$). Chiral exchange
alone (dotted) and an achiral bond in the same uniform field (blue dashed)
stay strictly on the isotropic locus (grey). Adding chirality to this uniform field ($\alpha=0.8\pi$, solid red) drives the propagator off the locus and directly through the CNOT
class (stars mark CNOT, iSWAP, and $\sqrt{\mathrm{SWAP}}$). The uniform field is
inert without chirality precisely because $\theta_{\mathrm{eff}}=\alpha$.
(b) The closest approach $d_{\min}$ (Euclidean distance in
$(\mathrm{Re}\,g_1,\mathrm{Im}\,g_1,g_2)$ over $Jt\le6\pi$) to the CNOT class
is plotted versus $\alpha$. The achiral uniform-field bound (blue dashed) and an explicit
orthogonal $90^{\circ}$ field gradient of the same magnitude (grey dotted) are
shown for comparison. Chirality reaches $d_{\min}=7\times10^{-4}$ at
$\alpha=0.85\pi$ and $Jt=1.01\pi$, which is closer than the explicit gradient
($d_{\min}=0.115$ at $Jt=4.70\pi$) and $4.7$ times faster.}
\label{fig:zeeman}
\end{figure*}

\fig{fig:zeeman}(a) plots the trajectories for the uniform-field geometry
corresponding to the second corollary. With chiral exchange alone, or with an achiral bond in the
uniform field, the propagator stays strictly on the isotropic locus. Adding chirality
to this uniform field drives the propagator off the locus and directly through the CNOT
class. Because the achiral uniform field remains completely inert
($\theta_{\mathrm{eff}}=0$ when $\alpha=0$), the entire departure from the
locus is solely attributable to chirality. This stands in contrast to an explicit field
gradient, which operates independently of the molecule.

The quantitative comparison is detailed in \fig{fig:zeeman}(b). A bare
isotropic bond approaches the CNOT class no closer than $d_{\min}=0.330$. An
explicit orthogonal gradient of magnitude $0.9J$ reaches $d_{\min}=0.115$ at
$Jt=4.70\pi$. The chiral synthetic gradient reaches $d_{\min}=7\times10^{-4}$
at $\alpha=0.85\pi$ in $Jt=1.01\pi$. This synthetic approach is both closer in invariant space and
nearly five times faster. The operational speed-up arises because the effective gradient
angle is generated by the fast exchange rotation rather than by a slow
Zeeman precession. Architecturally, this mechanism replaces a fabricated micromagnet or a
dedicated gradient electrode~\cite{Zajac2018,Noiri2022} with an
electrically silent property of the linker, whose sign is fully programmable by
enantiomer choice. This comparison uses equal field magnitudes and a single
free evolution as a deliberately conservative proxy. A fully compiled gate
sequence would improve these metrics further, but the single-evolution figure alone already
establishes the qualitative advantage.
\section{Spin shuttling as a non-Abelian holonomy}
\label{sec:holonomy}
Spin shuttling through a sequence of chiral bridges introduces a genuinely new and gauge-robust computational resource.

\subsection{Path-ordered holonomy}
Consider transporting a single electron through bridges
$1$ to $N-1$ with respective axes $\nv_1$ through $\nv_{N-1}$ and angles
$\alpha_1$ through $\alpha_{N-1}$. In the adiabatic and fully coherent limit
representing Scenario A, the spin acquires the path-ordered product
\begin{equation}
\Ush = U_{N-1}\,U_{N-2}\cdots U_2\,U_1
= \prod_{k=N-1}^{1}\exp\!\Big(\tfrac{i}{2}\alpha_k\,\nv_k\!\cdot\!\bm\sigma\Big).
\label{eq:holonomy}
\end{equation}
This mathematically constitutes a discrete non-Abelian holonomy representing an ordered product of $\SUtwo$
rotations about generally different spatial axes. Two operational regimes must be distinguished.

For collinear axes where $\nv_k\equiv\nv$, all matrix factors commute. The operator
$\Ush=\exp(\tfrac{i}{2}(\sum_k\alpha_k)\,\nv\!\cdot\!\bm\sigma)$ then collapses into a single
rotation about the common axis. If the spin is prepared and read along the
global field axis, and if $\nv\parallel\hat z$, this rotation is purely
diagonal and produces no observable spin flip. More generally, it is a fixed and path-independent rotation easily removable by a coordinate frame choice. This represents the Abelian and gauge-trivial case, serving as the transport analogue of the single-bond result.

For non-collinear axes, the matrix factors do not commute. The operator $\Ush$ depends intimately on
the specific order and physical geometry of the axes rather than merely on the sum
$\sum_k\alpha_k$. The resulting holonomy is genuinely non-Abelian and cannot be removed
by any single frame redefinition. This represents the operational regime of interest.

\subsection{Non-reciprocity}
\begin{figure*}[t]
\centering
\includegraphics[width=0.92\textwidth]{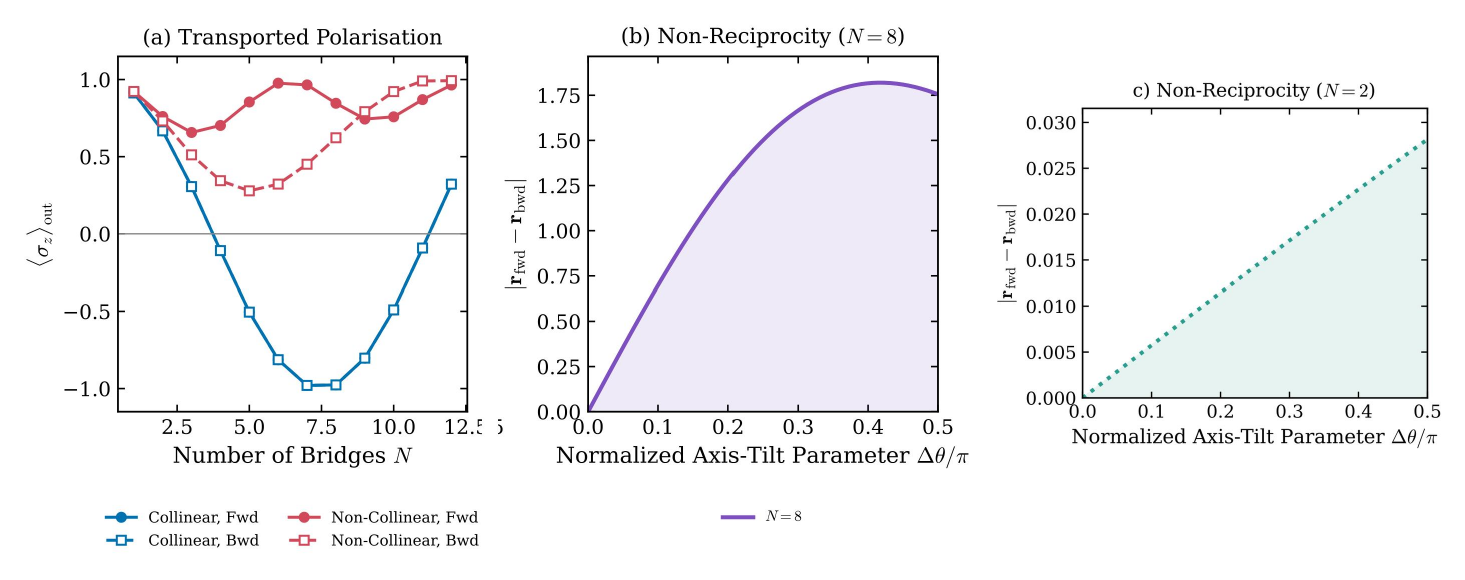}
\caption{Shuttling holonomy and non-reciprocity. 
(a) Transported polarisation $\langle\sigma_z\rangle_{\mathrm{out}}$ is plotted versus the number of bridges $N$ for an input polarised along $\hat z$. 
Collinear axes yield identical forward (circles) and backward (open squares) results, confirming they are reciprocal and gauge-trivial. 
Non-collinear axes yield distinct forward and backward outputs as the transport is rotationally non-reciprocal. 
(b) The non-reciprocity magnitude $\vert\rv_{\mathrm{fwd}}-\rv_{\mathrm{bwd}}\vert$ is evaluated versus the inter-bridge axis tilt for $N=8$. 
It vanishes exactly in the collinear limit and grows with geometric non-collinearity until it dips slightly below tilt = 0.5 on the x axis.}
\label{fig:holonomy}
\end{figure*}

Backward transport along a chain applies the inverted sequence
\begin{equation}
U_{\mathrm{sh}}^{\mathrm{rev}} = U_1^\dagger U_2^\dagger \cdots U_{N-1}^\dagger = U_{\mathrm{sh}}^\dagger,
\end{equation}
meaning a round trip returns the identity operation only if the spin is otherwise untouched. We also define the transport along the spatially inverted sequence of unitaries as $U_{\mathrm{sh}}$.
\begin{equation}
U_{\mathrm{sh}}^{\mathrm{bwd}} = U_1 U_2 \cdots U_{N-1} 
\end{equation}
The one-way transported state diverges between forward
and backward directions whenever the endpoint quantisation and readout axis is
fixed by the global magnetic field $\Bv_0$. Concretely, for an input polarised along
$\hat z$, the output polarisation $\langle\sigma_z\rangle$ after forward
versus backward shuttling differs strictly for non-collinear axes while coinciding perfectly for
collinear axes. \fig{fig:holonomy}(a) illustrates this effect. The collinear forward and backward curves perfectly overlap to demonstrate reciprocity, while the non-collinear curves separate.
\fig{fig:holonomy}(b) quantifies the specific non-reciprocity magnitude
$\vert\rv_{\mathrm{fwd}}-\rv_{\mathrm{bwd}}\vert$ as a function of the inter-bridge
axis tilt. This metric vanishes exactly at zero tilt and grows with
geometric non-collinearity. This directional dependence acts as a modest but true computational resource. It functions as an electrically silent and molecularly encoded diode-like action on the transported spin and serves as one of the critical experimental observables detailed in \sect{sec:channel}.

Two specific clarifications are necessary regarding this dynamic. First, the observables in
\fig{fig:holonomy} and in the closed-loop analysis below are computed in the
frame co-rotating with the global Larmor precession set by $\Bv_0$, which is mathematically equivalent to echoing out the deterministic Zeeman phase. In the laboratory frame, the finite continuous precession between bridges dresses each bridge axis with an arrival-time-dependent azimuth. We fold this variation into the axis-noise budget of \sect{sec:budget} rather than incorporating it into the geometric signal. Second, this rotational non-reciprocity represents a direction-dependent axis of the transported spin. This phenomenon is fundamentally distinct from the longitudinal non-reciprocity of an incoherent filter, which reverses the sign of the pumped polarisation (\sect{sec:channel}). The rotational non-reciprocity acts as a unique marker of coherence because the longitudinal filter asymmetry is present even for a fully classical spin device. Keeping these phenomena distinct is essential for the measurement programme proposed below.

\subsection{Closed-loop geometric gates}
The gauge-invariant physical content of the holonomy is best revealed by a closed
loop. Suppose the electron is shuttled outward along one arm designated by axes
$\{\nv_k^{\to}\}$ and returned along a second arm whose axes
$\{\nv_k^{\leftarrow}\}$ retrace a geometrically distinct path. The net
operation is a Wilson loop,
\begin{equation}
W = (\Ush^{\leftarrow})^{\dagger}\,\Ush^{\to} .
\end{equation}

Here \begin{equation}
\Ush^{\leftarrow} = V_1\,V_2\cdots V_{N-1},
\end{equation}
and 
\begin{equation}
\Ush^{\rightarrow} = U_{N-1}\,U_{N-2}\cdots U_{1},
\end{equation}

\begin{figure*}[t]
\centering
\includegraphics[width=0.92\textwidth]{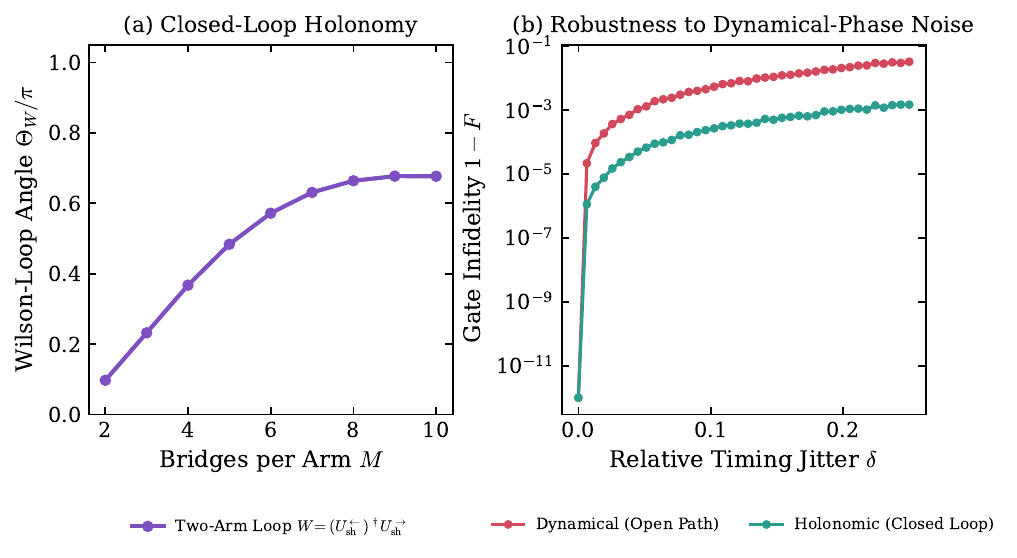}
\caption{Closed-loop geometric gate and its associated robustness. (a) The Wilson-loop
rotation angle $\Theta_W/\pi$ is plotted versus bridges per arm $M$ for a closed
shuttling loop. The angle is fixed by geometry and grows with loop size before
saturating. (b) The simulated gate infidelity is plotted versus relative timing jitter $\delta$. The
open-path dynamical gate (red) degrades linearly in $\delta$. The closed-loop
holonomic gate (green) effectively suppresses dynamical-phase noise to second order and is
far more robust at small $\delta$. This mathematical advantage narrows as $\delta$ grows and
does not extend to physical axis or path noise.}
\label{fig:geometric}
\end{figure*}

Here $\Ush^{\leftarrow}$ is the spatially inverted sequence of the unitaries
which coincides with $U_{\mathrm{sh}}$ if and only if the bridge rotations commute.
This loop is invariant under any local redefinition of the endpoint frame and therefore constitutes a
genuine physical single-qubit gate. \fig{fig:geometric}(a) plots the
Wilson-loop rotation angle $\Theta_W$ as a function of the number of bridges
per arm $M$. The angle grows with the loop size before saturating and is set purely by
the geometry of the two arms rather than by the traversal rate. An $8$-bridge
one-way path using the axis schedule of \fig{fig:holonomy} already accumulates
a net rotation of $0.732\pi$, confirming that substantial single-qubit
rotations are available purely during transport.

Because the Wilson-loop angle depends exclusively on the enclosed path in parameter space and not
on the traversal speed, closed-loop shuttling gates inherit the established noise
resilience of holonomic quantum
computation~\cite{Zanardi1999,Duan2001,Sjoqvist2015}. \fig{fig:geometric}(b)
quantifies this resilience specifically for dynamical-phase (timing) noise. We model a
fluctuating dynamical phase $\theta_k=\omega\tau_k$ acquired per bridge due to
local Zeeman precession during a fluctuating dwell time $\tau_k$. The
open-path dynamical gate degrades linearly in the timing jitter $\delta$. Conversely, the closed-loop holonomic gate echoes away the forward-arm dynamical phase on the return arm, causing it to degrade only quadratically to second order. The holonomic gate is therefore markedly more robust at small
noise amplitudes. We rigorously delimit this claim. The second-order suppression is specific to dynamical-phase and timing noise. Fluctuations in the physical axes
$\nv_k$ or the geometry of the path due to charge-noise-induced changes in molecular conformation perturb both gate types at first order and are
not suppressed by the geometric construction. Holonomic robustness constitutes a targeted rather than a universal advantage.

\subsection{Relation to two-terminal reciprocity}
\label{sec:onsager}
It is critical to distinguish this rotational non-reciprocity from constraints governing two-terminal charge transport. For two-terminal charge transport in the linear-response regime,
Onsager-Casimir symmetry forces the overall conductance to be even in a magnetic
field regardless of the presence of strong spin-orbit coupling. Consequently, a genuine
two-terminal CISS magnetoresistance at $B=0$ is theoretically forbidden~\cite{Yang2019,Evers2022}. Our claim lies completely outside the scope of that limitation on three distinct counts. First, the observable is the quantum spin state of a single
transported electron rather than a linear-response steady charge current. Second, it is read
out projectively at a fixed working field $\Bv_0\neq0$. Third, it is conditioned
on a successful and essentially unidirectional single-electron shuttling event rather than
averaged over a continuous two-terminal thermodynamic flow. The rotational non-reciprocity of
\fig{fig:holonomy} is therefore a targeted statement about coherent single-spin
transport. It is not a violation of the charge-conductance symmetry and should
not be mathematically conflated with one. The longitudinal filtering asymmetry explored in \sect{sec:channel} is likewise a highly nonlinear and state-dependent quantum effect rather than a linear-response conductance parameter.

\subsection{Role of the holonomy in two-qubit circuits}
\label{sec:vd}
We now define the exact operational role of the shuttling holonomy when a mobile qubit is delivered into exchange contact with a
stationary qubit. The holonomy $\Ush$ acts as a strictly local unitary on the
transported qubit. Following the rigorous arguments of \sect{sec:gauge}, it cannot change the
two-qubit local-equivalence class. This core class is fixed entirely by the final exchange
segment alongside non-collinear fields when an operation beyond the SWAP
family is required (\sect{sec:zeeman}). The role of the holonomy is therefore compilatory
rather than class-generating. Any target two-qubit unitary rigorously factorises as
\begin{equation}
U_{\mathrm{target}}=(k_1\otimes k_2)\,U_{\mathrm{nonlocal}}\,(k_3\otimes k_4),
\end{equation}
utilising four surrounding single-qubit gates $k_1$ through $k_4$. Of these components, the two
acting on the mobile qubit ($k_2$ and $k_4$ executed before and after the
exchange contact) are exactly the operations the transport legs supply intrinsically. These operations are executed concurrently with the baseline spatial motion that the architecture
performs regardless. A shuttling-based two-qubit gate thereby absorbs two of the
four flanking local rotations directly into the transport sequence at no additional gate time,
provided the bridges remain sufficiently coherent for these rotations to be unitary (with the required tolerance for $\Fc\to1$ quantified in \sect{sec:budget}). The
distinctive and gauge-invariant contributions of chirality remain the geometric
single-qubit gate acquired during transport and the rotational
non-reciprocity, both of which are absent for collinear or achiral bridges. The entangling
nonlocality itself arises from the targeted exchange interaction as established in \sect{sec:zeeman}.

\section{Measuring the coherent fraction and the experimental programme}
\label{sec:channel}
The theoretical results of \sect{sec:holonomy} inherently assume fully coherent Scenario A bridges.
Whether a real fabricated chiral bridge acts coherently or operates as an incoherent spin filter resembling Scenario B remains a decisive experimental question for device
viability because only the coherent limit supports holonomic gates. We demonstrate
that the specific channel defined in \eq{eq:channel}, and in particular the coherent fraction
$\Fc$, is rigorously over-determined by a small set of transport measurements governed by the exact
composition laws derived previously in \sect{sec:unified}.

\subsection{Polarisation pumping and transverse decay}
\begin{figure*}[t]
\centering
\includegraphics[width=0.92\textwidth]{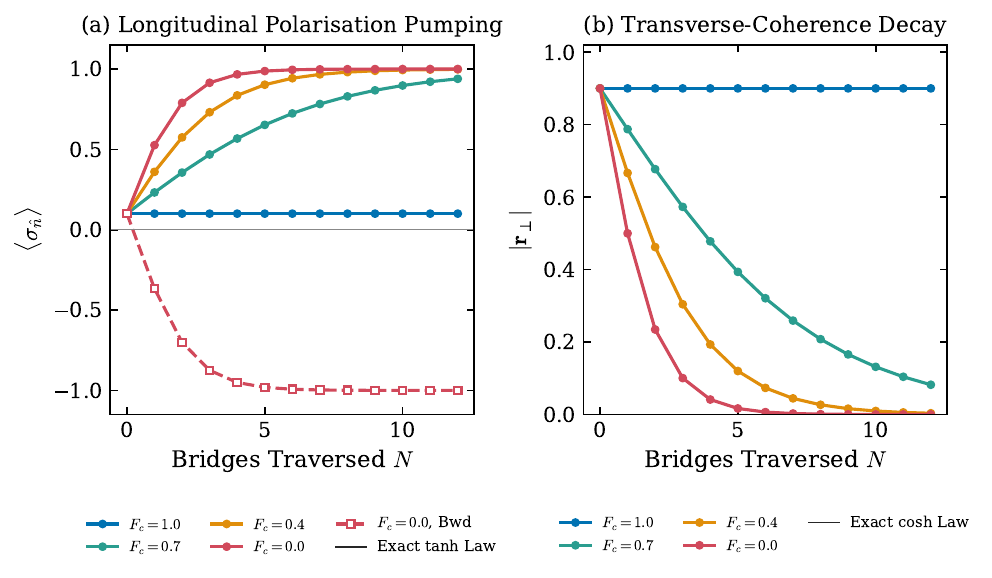}
\caption{The filtering and dephasing bridge channel. (a) The longitudinal
polarisation $\langle\sigma_{\nv}\rangle$ of the transported spin is plotted versus
bridge number $N$ for coherent fractions $\Fc=1.0,0.7,0.4,0.0$
using parameters $\lambda_f=0.45$, $\lambda_d=0.35$, and common axis $\nv$. Forward traversal
(filled markers) pumps toward $+1$. The simulated $\Fc=0$ chain run in reverse (open
squares) pumps toward $-1$, clearly exhibiting the direction reversal of the filter.
The solid black curve is the exact theoretical hyperbolic tangent law from \eq{eq:tanh}. (b) Transverse
coherence $\vert\rv_\perp\vert$ is plotted versus $N$, where the solid curve maps the closed-form
hyperbolic cosine ratio law from \eq{eq:perpN}. Because the pure states $\pm\nv$ are fixed points of the
filter, state purity is non-monotonic and is not a valid coherence
witness. The longitudinal pump and transverse decay remain the rigorous metrics.}
\label{fig:channel}
\end{figure*}

Two families of simulated curves carry this physical information through exact closed-form laws.
The longitudinal polarisation of the transported spin obeys the hyperbolic tangent
composition law from \eq{eq:tanh}. An initially unpolarised or weakly polarised
input is pumped steadily toward $\pm\nv$ with a designated rapidity of $a=\artanh\kappa$ per bridge.
Because the underlying filter sign tracks the propagation direction, the longitudinal pump
reverses under $s\to-s$. \fig{fig:channel}(a) illustrates this specific pumping dynamic for several
values of $\Fc$ alongside the direction-reversed $\Fc=0$ curve. The solid line represents the
exact analytical law, which the simulated discrete channel reproduces to a precision of $2\times10^{-16}$. The
transverse coherence obeys \eq{eq:perpN}, which forms a product of a geometric dephasing
factor $(1-q)^N$ and a purely spatial hyperbolic cosine ratio. \fig{fig:channel}(b) plots
this monotone transverse decay, which is matched by the theoretical closed form.

The filter's fixed points introduce a counterintuitive dynamic that precludes using state purity as a direct experimental metric. Because the filter's fixed points are the pure states $\pm\nv$, the physical purity of the transported spin remains non-monotonic in bridge number and returns toward unity as the entire chain
polarises. Purity is therefore fundamentally not a reliable witness of physical incoherence. An
experiment tracking only $\Trace\rho^2$ would mathematically misdiagnose a strong
classical filter as a highly coherent channel. The correct experimental observables are the
longitudinal pump to directly measure $\kappa$ and the transverse
decay at fixed longitudinal polarisation to uniquely isolate $q$. One single traversal of an unpolarised input yields $\rpar^{(1)}=s\kappa$, demonstrating that $\kappa$ represents exactly the macroscopic transmission polarisation reported in standard CISS
experiments and remains the sole parameter currently constrained by existing literature.

\subsection{Four complementary observables}
Collecting the coherent geometric signatures of \sect{sec:holonomy} alongside the filtering
signatures yields four distinct observables that together  over-determine $\Fc$.
\begin{enumerate}
\item[$O_1$] \emph{Rotational non-reciprocity} [\fig{fig:holonomy}]. This provides a
direction-dependent transported \emph{axis} generated from a non-collinear coherent
chain. It is present only for $\Fc>0$ acting coherently because a classical incoherent filter cannot
produce a coherent axis rotation. This makes it a rigorous signature of coherence.
\item[$O_2$] \emph{Closed-loop Wilson interference} [\fig{fig:geometric}]. This represents a
nontrivial geometry-determined unitary $W$, determined purely by interfering the
returned spin with a stable reference. It is present only in the coherent limit and serves as a secondary coherence witness.
\item[$O_3$] \emph{Longitudinal pumping} [\fig{fig:channel}(a)]. This follows the exact hyperbolic tangent law
and is direction-reversing. The per-bridge rapidity gives the metric $\kappa=(1-\Fc)\lambda_f$.
It is present only for $\Fc<1$. Its exact sign reversal under path reversal structurally distinguishes it from $O_1$.
\item[$O_4$] \emph{Transverse decay} [\fig{fig:channel}(b)]. This obeys the hyperbolic cosine ratio
law where the specific dephasing factor provides $q=(1-\Fc)\lambda_d$ once $\kappa$ is physically known
from $O_3$.
\end{enumerate}
The pair $(O_3,O_4)$ yields two fully independent estimates of the target coherent
fraction, specifically $\widehat{\Fc}=1-\widehat{\kappa}/\lambda_f$ and
$\widehat{\Fc}=1-\widehat{q}/\lambda_d$. The parameters $\lambda_f$ and $\lambda_d$ are fixed
by a racemic, achiral, or high-temperature calibration (detailed in the Supplemental Material).
Their mathematical agreement provides a rigorous internal consistency test. Furthermore, any underlying coherent
contribution additionally activates $O_1$ and $O_2$, which the classical statistical filter
cannot replicate. Because no single scalar classical mechanism reproduces all four metrics, the
discrimination is highly robust against the standard loopholes that complicate two-terminal
CISS magnetoresistance measurements~\cite{Evers2022}. The critical delimitation of
\sect{sec:onsager} applies to $O_3$ specifically, as it is a nonlinear and state-dependent pumping effect rather than a restricted linear-response conductance.

\section{Silicon-specific physics and the valley degree of freedom}
\label{sec:valley}
A realistic \Sisotope\ implementation must account for the valley degree of freedom, which has no mathematical analogue in the standard molecular-spintronics setting but remains
decisive for operational silicon spin
qubits~\cite{Zwanenburg2013,Friesen2007,Culcer2010}. The bulk sixfold valley
degeneracy is physically lifted to a twofold degeneracy by device confinement. The residual
two low-lying valleys are further split by an interface-dependent valley
splitting parameter $E_v$. This splitting is typically $E_v\sim20$ to $300\,\mu$eV and remains highly sensitive to atomic-scale interface disorder.

Two specific valley effects have a direct influence on our proposal. First, coherent spin transport
requires that shuttling remain mathematically adiabatic with respect to the valley
gap. If the electron traverses a bridge or interdot region faster than the timescale set by
$\hbar/E_v$, or if the localized valley composition varies sharply along the path, intervalley
transitions will leak population out of the defined computational subspace and effectively randomise
the spin through induced spin-valley mixing. Second, the spin
transformation of the chiral bridge from \eq{eq:Uij} is defined strictly within a single valley subspace. Strong
valley mixing at the physical molecule-silicon interface would rapidly average over valleys
featuring different effective spin-orbit couplings and forcefully degrade the coherent
holonomy toward the incoherent limit associated with a small $\Fc$.

\begin{figure*}[t]
\centering
\includegraphics[width=0.92\textwidth]{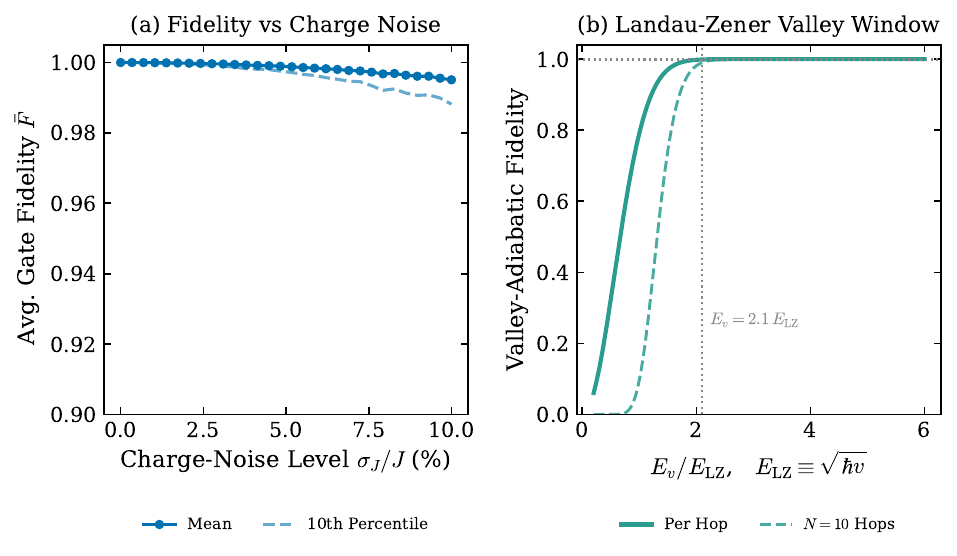}
\caption{System error budget. (a) The average gate fidelity $\bar F$ of a
shuttling-assisted two-qubit operation is plotted versus the quasistatic charge-noise level
$\sigma_J/J$. The plot details the mean (solid) and $10$th percentile (dashed) over thousands of simulated
noise realisations. The barrier-sensitive rotation angle is co-fluctuated at
$\sigma_\alpha/\alpha=\sigma_J/2J$ as detailed in \sect{sec:budget}. (b) The Landau-Zener
valley-adiabatic fidelity per hop (solid) and combined over ten hops (dashed) is plotted versus
$E_v/\ELZ$, where $\ELZ\equiv\sqrt{\hbar v}$. The vertical dotted lines mark the precise
$E_v=2.1\,\ELZ$ threshold required for $99.9\%$ per-hop adiabaticity. The simulated charge noise is
modelled conservatively as quasistatic white noise rather than realistic $1/f$ noise as noted in \sect{sec:limitations}.}
\label{fig:fidelity}
\end{figure*}

The first effect constitutes a genuine Landau-Zener problem, and we treat it as one
rather than utilizing a soft interpolation. As the electron is shuttled, the applied interdot
detuning sweeps the lower valley through its physical anticrossing with the upper
valley. Modelling the sweep as completely linear at rate $v=\vert\dot\varepsilon\vert$ through a localized
gap $E_v$ yields a diabatic leakage probability per hop of
\begin{equation}
P_{\mathrm{LZ}}=\exp\!\Big(-\frac{\pi E_v^{2}}{2\hbar v}\Big).
\label{eq:lz}
\end{equation}
Consequently, maintaining valley-adiabatic transport requires $E_v\gtrsim2.1\,\ELZ$ with the energy scale defined as
$\ELZ\equiv\sqrt{\hbar v}$. The specific numerical factor enforces a strict threshold of
$P_{\mathrm{LZ}}\le10^{-3}$ per hop. \fig{fig:fidelity}(b) plots the
per-hop and combined ten-hop valley-adiabatic fidelities against the normalized parameter $E_v/\ELZ$. This
criterion dictates the required detuning sweep rate rather than the internal tunnel coupling. Faster shuttling directly raises $\ELZ$ and inherently demands a significantly larger valley splitting to prevent leakage. For a standard detuning excursion of a few meV completed in $1\,$ns, the energy scale is
$\ELZ\!\approx\!36\,\mu$eV and the required operational window is $E_v\gtrsim75\,\mu$eV. Slowing the identical excursion to $10\,$ns lowers $\ELZ$ to roughly $11.5\,\mu$eV and considerably relaxes the
window to $E_v\gtrsim24\,\mu$eV. This dynamic exposes a critical speed-adiabaticity
tension entirely unique to silicon devices. Fast shuttling mitigates ambient spin dephasing but severely exacerbates intervalley leakage, meaning the two opposing rates must be strictly balanced against the measured $E_v$ of a given physical interface. A full theoretical treatment incorporating a
position-dependent valley phase and realistic atomic disorder is provided by Langrock
\emph{et al.}~\cite{Langrock2023}. Equation (\ref{eq:lz}) establishes the explicit leading-order operational window within which their theoretical refinements apply.

\section{Error budget and feasibility constraints}
\label{sec:budget}
We estimate the leading error channels using strictly conservative parameter bounds to establish system feasibility.

The dominant decoherence source for quantum exchange and
shuttling operations in silicon remains electrical charge noise on the confinement gates, which
modulates $t$ and consequently fluctuates $J=4t^2/U_H$~\cite{Yoneda2018,Nakajima2020}. Because the scaling is
$J\propto t^2$, the final fractional exchange fluctuation is exactly twice the fractional
tunnelling fluctuation. \fig{fig:fidelity}(a) plots the resulting average gate fidelity
of a shuttling-assisted two-qubit operation as a direct function of the quasistatic
charge-noise level $\sigma_J/J$. Maintaining operational fidelities above $99\%$ fundamentally requires
$\sigma_J/J\lesssim3\%$, which is mathematically consistent with the charge-noise levels reported in
state-of-the-art \Sisotope/SiGe semiconductor devices. The electrostatic barrier setting $t$ simultaneously sets
the bridge rotation angle $\alpha$, forcing us to safely co-fluctuate the two parameters. Because
$J\propto t^2$ is quadratic while $\alpha$ scales linearly with the
barrier-controlled coupling, we apply the relation $\sigma_\alpha/\alpha=\sigma_t/t=\sigma_J/2J$.
This specific coupled fluctuation has a truly negligible effect on the final error budget. Freezing $\alpha$ completely while
retaining the $J$ noise alters the mean simulated fidelity by less than $5\times10^{-4}$ at
$\sigma_J/J=6\%$ because the overall fidelity is strictly dominated by the primary exchange-angle
error. This particular estimate assumes a highly simplified quasistatic white noise
model. Real laboratory devices exhibit complex $1/f$ spectra featuring slow components that can be partly
echoed out, though their cumulative effect on extended multi-bridge shuttling sequences is substantially more severe than what the
white-noise baseline suggests (\sect{sec:limitations}).

Regarding valley leakage, maintaining the strict Landau-Zener window of $E_v\gtrsim2.1\sqrt{\hbar v}$ established in \sect{sec:valley} effectively suppresses diabatic transitions. Because this energy window is set entirely by the detuning sweep
rate $v$ rather than the internal tunnel coupling, transport fidelity falls precipitously when operational parameters exit this regime.

Bridge-induced decoherence introduces a separate and highly stringent constraint. If the molecular bridge is even partially incoherent, featuring $\Fc<1$, each spatial traversal multiplies the transverse quantum coherence by
$(1-q)\sqrt{1-\kappa^{2}}$ according to \eq{eq:transverse}. To first order in the filter strength, this mechanism additionally deforms the transported state through the longitudinal polarisation pump. For an entire chain of $N$ bridges, the accumulated transverse loss scales exactly as
$1-[(1-q)\sqrt{1-\kappa^2}]^{N}$. Keeping this cumulative loss below a $1\%$ threshold over a sequence of $N\sim10$ operations fundamentally requires $\max(\lambda_f,\lambda_d)(1-\Fc)\lesssim10^{-3}$. This bounds the coherent fraction to
$\Fc\gtrsim1-10^{-3}/\max(\lambda_f,\lambda_d)$, which must be exceedingly close to unity. This is the most difficult physical requirement of the device proposal and represents the parameter least
constrained by existing empirical data. Current experiments mathematically fix only the combined product $\kappa=(1-\Fc)
\lambda_f$. Consequently, whether complex molecular bridges can act sufficiently coherent at the critical single-electron and cryogenic level remains unknown at present.

The resulting theoretical scheme is therefore not intended as a near-term device blueprint. It's fundamentally conceptual and diagnostic. It isolates the exact geometric holonomy, the
rotational non-reciprocity, and the targeted synthetic gradient as the only genuine chiral
resources. It additionally provides the four complementary observables of \sect{sec:channel} required to physically measure $\Fc$ and establishes rigorous quantitative performance targets. Candidate chiral-bridge systems must ultimately be judged strictly against the required physical bounds $E_v\gtrsim2.1\sqrt{\hbar
v}$, $\sigma_J/J\lesssim3\%$, and $\Fc\gtrsim1-10^{-3}/\max(\lambda_f,\lambda_d)$.

\section{Approximations and underlying limitations}
\label{sec:limitations}
We present the explicit assumptions upon which these mathematical results depend to ensure they are fully contextualized outside the main derivations.

The twisted-exchange Hamiltonian derived in \eq{eq:twisted} originates from a second-order Schrieffer-Wolff reduction and represents the absolute leading term scaling as $t^2/U_H$. Fourth-order corrections including generalized ring exchange alongside adjustments to $J$ and the internal anisotropy enter the system at exactly $O(t^4/U_H^3)$. These higher-order terms are mathematically estimated to shift the effective couplings by $5$ to $10\%$ in the typical operating regime characterized by $t/U_H\sim0.1$. The underlying gauge-triviality result remains fully exact at each perturbative order, meaning any purely bond-local DM interaction is fully removable, and higher-order corrections do not alter this fundamental mathematical conclusion.

The shuttling holonomy operator derived in \eq{eq:holonomy} assumes the transported electron follows its instantaneous orbital and valley ground state. Diabatic operational corrections enter the system directly as valley Landau-Zener leakage, which is controlled by the energy window
$E_v\gtrsim2.1\sqrt{\hbar v}$ defined in \eq{eq:lz}. The resulting operational speed-adiabaticity tension is quantified in \sect{sec:valley}. We work strictly within the subspace of one orbital and one valley. Orbital excitation during fast dynamic shuttling and severe valley mixing at the physical interface are treated only phenomenologically through $\Fc$ and the integrated leakage estimate.

The physical parameters defining the channel triplet $(\Fc,\lambda_f,\lambda_d)$ are phenomenological and are not derived from a dedicated microscopic transport calculation. A first-principles or NEGF treatment of the bridge structure that outputs these specific parameters as mathematical functions of the
molecule, temperature, and applied bias remains the most critical missing component and is
left for future theoretical work. The specific rotation angles $\alpha$
required for observable geometric holonomy, rotational non-reciprocity, and the targeted synthetic gradient
occupy the high optimistic end of what realistic chiral bridges might theoretically deliver. Unlike
the filtering polarisation $\kappa$ which existing CISS data constrain, the specific parameter $\alpha$ remains
experimentally unconstrained at present. No currently reported measurement formally fixes it, and properly determining it remains one primary purpose of the specific protocols proposed here. Reported CISS physical magnitudes and even core sign conventions
remain intensely debated in the literature~\cite{Evers2022,Fransson2020,Alwan2021}, and we have intentionally not
assumed a highly specific microscopic CISS mechanism.

The filtering channel mathematically defined in \eq{eq:channel} is conditioned purely on a successful physical shuttling event. A failed charge transfer constitutes a
heralded physical error detectable by external charge sensing rather than a purely quantum spin
error. The fully unconditional mathematical map therefore includes a classical leakage and loss branch, not represented within the idealized Bloch-sphere laws. The theoretical polarisation pump $O_3$ is therefore strictly the transfer-heralded expectation value. This structural reality is highly favourable for metrology using herald-and-postselect techniques but must be accounted for in any comprehensive fidelity accounting
of a fully working logical gate.

The formal error budget utilizes a simplified quasistatic white
noise spectrum. Realistic $1/f$ noise spectra will modify the theoretical fidelity estimates, generally operating unfavourably for long transport chains, although holonomic echoing protocols naturally mitigate the slowest
frequency components. No actual chiral-bridge silicon quantum-dot device matching the parameters of the kind analysed here has been physically fabricated. The theoretical proposal functions as a mathematical target, and the quantitative physical windows established above are exact predictions to be tested rather than retroactively fitted results.

\section{Discussion and formal conclusion}
\label{sec:conclusion}
We have systematically analysed spin shuttling through chiral molecular bridges in silicon
quantum-dot chains to establish a rigorous operational boundary separating what chirality can and cannot computationally provide. A single chiral bond does not constitute a new two-qubit resource, regardless of the magnitude of its CISS-induced
spin rotation. The twisted exchange it mathematically generates remains locally equivalent to isotropic Heisenberg exchange. We confirmed this gauge symmetry with exact precision using the Makhlin invariants and traced the equivalence directly to the inescapable symmetric anisotropy that accompanies any local bond
Dzyaloshinskii-Moriya term. The common theoretical proposals for continuously tunable chiral gates rest entirely upon the mathematically inconsistent truncation of that exact symmetric term.

Genuine computational resources emerge only when the single-bond gauge
freedom is fundamentally broken. The gauge analysis identifies precisely when this occurs, demonstrating that the two-qubit class depends on the bridge angle exclusively through the relative field
geometry $\angle(\Bv_1,\Rmat_\alpha^{-1}\Bv_2)$. This criterion yields a highly
constructive corollary. Chiral exchange turns a uniform transverse
field into a synthetic gradient featuring a chemically programmable angle. This gradient approaches the CNOT class more closely and operates nearly five times faster than an explicit physical gradient of equal magnitude, effectively replacing a fabricated macroscopic micromagnet with an intrinsic property of the molecular linker. 

The most distinctive and robust chiral resource is the shuttling holonomy. Transporting a spin through non-collinear chiral bridges realises a genuine non-Abelian $\SUtwo$ holonomy exhibiting rotational
non-reciprocity. In a closed-loop form, this implements a geometry-determined
single-qubit gate that naturally suppresses dynamical-phase noise exactly to second order. Within a larger two-qubit circuit, its operational role is to compile the flanking local rotations of the mobile qubit directly into the transport sequence itself. 

We treated the coherent and incoherent
interpretations of CISS on an equal mathematical footing using a continuous coherent fraction $\Fc$. By modelling the incoherent branch as a direction-reversing spin filter, we derived exact composition laws including a hyperbolic tangent law for the pumped polarisation and a closed-form transverse-decay law. These laws yield four
complementary physical observables that strictly over-determine $\Fc$. This framework ties the theoretical filtering
parameter $\kappa$ directly to the empirical transmission polarisation reported by existing CISS
experiments.

We explicitly note that this theoretical characterisation does not constitute a blueprint for a near-term working device. The
coherence requirement imposed on the bridge is extremely severe and remains presently unconstrained in the critical single-electron
cryogenic regime. The specific rotation angle $\alpha$ required to manifest coherent effects is itself unmeasured. Furthermore, silicon valley physics
imposes a strict Landau-Zener adiabaticity window of $E_v\gtrsim2.1\sqrt{\hbar v}$ that directly limits the maximum permissible shuttling speed. The primary contribution of this work is to replace qualitative expectations with a precise theoretical boundary. We separate the gauge artifact
from the genuine chiral resources, derive the invariants and exact channel laws required to certify them, and formally specify the measurements required to decide whether
chiral bridges constitute a viable physical addition to the silicon spin-qubit architecture. The most
valuable next step is a detailed microscopic transport calculation capable of predicting
$(\Fc,\lambda_f,\lambda_d)$ for specific chiral linkers attached to
hydrogen-terminated silicon, which can successfully convert this phenomenological mathematical channel into physical material parameters.

\begin{acknowledgments}
We gratefully acknowledge financial support from the National Quantum Mission (NQM) of the Department of Science and Technology (DST), Government of India, through Grant No. DST/QTC/NQM/QC/2024/1. We also acknowledge the use of Claude (Anthropic) for assistance with literature review and manuscript preparation.
\end{acknowledgments}

\appendix

\section{Schrieffer-Wolff derivation of the twisted exchange}
\label{app:sw}
Consider two dots at half filling defined by the spin-dependent hopping
in \eq{eq:hop} and on-site interaction $U_H$. In the singly-occupied low-energy subspace, virtual hops to a doubly-occupied
dot and back generate an effective spin coupling. We write the hopping as
$T=-t\sum_{\sigma\sigma'} c^{\dagger}_{1\sigma}(U)_{\sigma\sigma'}c_{2\sigma'} +\text{h.c.}$ and place the doubly-occupied manifold at energy $U_H$. The
standard second-order Schrieffer-Wolff transformation~\cite{Bravyi2011} yields
$\Heff = -\,P_0\,T\,(H_0-U_H)^{-1}\,T\,P_0$ with $P_0$ acting as the projector onto the
singly-occupied subspace. Evaluating the two virtual processes mapping $1\to2$
and back alongside $2\to1$ and back reveals that the spin structure carried by the
$\SUtwo$ matrix $U$ mathematically produces
\begin{equation}
\Heff=\frac{4t^2}{U_H}\;\sum_{ab}\,(\Rmat)_{ab}\,S_1^{a}S_2^{b}
= J\,\Sv_1\!\cdot\!(\Rmat\Sv_2),
\end{equation}
where the defining matrix $\Rmat$ satisfies
$U(\bm a\!\cdot\!\bm\sigma)U^{\dagger}=(\Rmat\bm a)\!\cdot\!\bm\sigma$ and represents
the $\SOthree$ image of $U\in\SUtwo$. For the specific transformation
$U=\exp(\tfrac{i}{2}\alpha\,\nv\!\cdot\!\bm\sigma)$, we evaluate
$\Rmat=\exp(\alpha\,\nv\!\cdot\!\bm L)$. Expanding this term using the Rodrigues
identity $\Rmat = \mathbb{1}+\sin\alpha\,[\nv]_\times +(1-\cos\alpha)[\nv]_\times^2$ yields the exact decomposition in \eq{eq:decomp}.
Fourth-order terms contribute at $O(t^4/U_H^3)$ and are neglected. Although these effects renormalize the specific coefficients, they remain fully expressible in terms of bond rotations and therefore do not compromise the fundamental gauge argument presented in Appendix~\ref{app:gauge}.

\section{Explicit gauge transformation}
\label{app:gauge}
We define the local unitary $V_2=\exp(-\tfrac{i}{2}\alpha\,\nv\!\cdot\!\bm\sigma)$ acting exclusively on spin $2$. Under conjugation, the internal spin operators transform as a vector such that
$V_2^{\dagger}S_2^{b}V_2=\sum_c (\Rmat^{-1})_{bc}\,S_2^{c}$. This occurs because
$V_2^\dagger$ physically implements the inverse $\SOthree$ rotation $\Rmat^{-1}$ on the
site-2 spin. Substituting this result into the twisted exchange yields
\begin{align}
V_2^{\dagger}\Heff V_2
&= J\sum_{abc}(\Rmat)_{ab}(\Rmat^{-1})_{bc}\,S_1^{a}S_2^{c}
\nonumber\\
&= J\sum_{ac}\delta_{ac}\,S_1^{a}S_2^{c}
= J\,\Sv_1\!\cdot\!\Sv_2 ,
\end{align}
utilizing the identity $\Rmat\Rmat^{-1}=\mathbb{1}$. The twist is therefore mathematically removed exactly by
a local unitary on a single spin, rendering $\Heff$ local-unitary equivalent to the
isotropic Heisenberg exchange. The identical operation $V_2$ rotates any site-2 Zeeman field according to
$\Bv_2\to\Rmat^{-1}\Bv_2$. This field rotation is undone only if $\Bv_2\parallel\nv$, leaving the isotropic exchange
intact. This specific geometric requirement demonstrates why a non-collinear field
in \sect{sec:zeeman} cannot be gauged away and restores genuine nonlocality.

\section{Makhlin local invariants}
\label{app:makhlin}
For a chosen two-qubit gate $U\in U(4)$, we transform the system to the magic Bell basis via
$U_B = Q^\dagger U Q$, utilizing the specific matrix
\begin{equation}
Q=\frac{1}{\sqrt2}\begin{pmatrix}
1&0&0&i\\ 0&i&1&0\\ 0&i&-1&0\\ 1&0&0&-i
\end{pmatrix}.
\end{equation}
We define the intermediate matrices $m=U_B^{T}U_B$ and $\bar U=U/(\det U)^{1/4}$. The resulting Makhlin invariants
are derived as
\begin{equation}
g_1=\frac{\big(\Trace\,\bar m\big)^2}{16},\qquad
g_2=\frac{(\Trace\,\bar m)^2-\Trace(\bar m^2)}{4},
\end{equation}
with $\bar m$ formed directly from $\bar U$. These specific metrics are invariant under the transformation
$U\to(u_1\otimes u_2)U(v_1\otimes v_2)$ and therefore remain constant on each
local-equivalence class. Two distinct gates mathematically share $(g_1,g_2)$ if and only if they are
local-unitary equivalent. For isotropic exchange, the physical trajectory
$U(t)=\exp(-iJ\,\Sv_1\!\cdot\!\Sv_2\,t)$ traces the well-known SWAP family with $g_2$ behaving as illustrated in
\fig{fig:gauge}(b). The mathematical identity of the chiral-bond and isotropic curves in
that figure constitutes the numerical verification of Appendix~\ref{app:gauge}. Reference
values utilized as landmarks in \fig{fig:zeeman} are computed directly from the exact gate
matrices in the companion code. These specific coordinates map to $(\mathrm{Re}\,g_1,g_2)=(0,+1)$ for
CNOT/CZ, $(0,-1)$ for iSWAP, $\mathrm{Re}\,g_1=0$ with $\mathrm{Im}\,g_1=-\tfrac14$ and
$g_2=0$ for $\sqrt{\mathrm{SWAP}}$, and $(-1,-3)$ for full SWAP. Mapping these static invariant coordinates directly to the dynamical trajectories shown in \fig{fig:zeeman}(a) confirms the exact topological intersection of the synthetic gradient propagator with the established CNOT class. The theoretical entangling
power quoted in the main text follows the exact closed form $e_p(U)=\tfrac29(1-\vert{}g_1 \vert{})$~\cite{Zanardi2000,Makhlin2002}. Because this metric depends strictly on the local invariant
$g_1$, it is mathematically constant on each local-equivalence class. It evaluates
to $1/6$ at the $\sqrt{\mathrm{SWAP}}$ point where $\vert{}g_1\vert{}=1/4$, precisely $2/9$ for the CNOT gate where
$g_1=0$, and $0$ for the SWAP operation where $\vert{}g_1\vert{}=1$.

\section{Holonomy, ordering, and non-reciprocity}
\label{app:holonomy}
The shuttling operator defined in \eq{eq:holonomy} is a path-ordered product. Evaluating the system for two
bridges yields
\begin{equation}
\begin{split}
\Ush = U_2 U_1 ={}&
\Big(\cos\tfrac{\alpha_2}{2}
     +i\sin\tfrac{\alpha_2}{2}\,\nv_2\!\cdot\!\bm\sigma\Big)\\
&\times\Big(\cos\tfrac{\alpha_1}{2}
     +i\sin\tfrac{\alpha_1}{2}\,\nv_1\!\cdot\!\bm\sigma\Big).
\end{split}
\end{equation}
Applying the Pauli vector identity
\begin{equation}
(\bm a\!\cdot\!\bm\sigma)(\bm b\!\cdot\!\bm\sigma)
=\bm a\!\cdot\!\bm b+i(\bm a\times\bm b)\!\cdot\!\bm\sigma
\end{equation}
reveals that the matrix product contains a specific term
$\propto(\nv_2\times\nv_1)\!\cdot\!\bm\sigma$. This mathematical term is antisymmetric under the swap
$1\leftrightarrow2$ and vanishes only under the condition $\nv_1\parallel\nv_2$. Consequently, the relation
$U_2U_1\neq U_1U_2$ holds rigorously for non-collinear axes, proving that the net rotation depends closely on the
order in which the physical bridges are traversed. 
Reversing the specific shuttling direction applies the inverted sequence $U_{\mathrm{sh}}^{\mathrm{rev}} = \Ush^\dagger = U_1^\dagger U_2^\dagger$. The forward and backward one-way outputs, when referred directly to the fixed laboratory axis set by $\Bv_0$, must therefore theoretically differ. The specific non-reciprocity magnitude plotted in \fig{fig:holonomy}(b) maps to $|\rv_{\mathrm{fwd}} - \rv_{\mathrm{bwd}}|$, utilizing the relation {$\rv_{\mathrm{fwd/bwd}} = \mathrm{Tr}[\bm{\sigma}\,\Ush^{\mathrm{fwd/bwd}}\rho_0(\Ush^{\mathrm{fwd/bwd}})^\dagger]$}. This metric vanishes identically in the pure collinear limit and
grows alongside the inter-bridge tilt dictates for the $N=2$. A closed loop directed forward along one arm and
returned along a geometrically distinct arm yields the gauge-invariant Wilson
loop $W$ discussed in \sect{sec:holonomy}. The specific rotation angle
$\Theta_W=2\arccos(\vert{}\Trace W\vert{}/2)$ is physically independent of any chosen endpoint frame.

\section{Bloch-vector recursion for the bridge channel}
\label{app:channel}
We explicitly derive the exact mathematical laws presented in \eqref{eq:mobius} through \eqref{eq:perpN}. We formalize the state density matrix as
$\rho=\tfrac12(\mathbb 1+\rv\!\cdot\!\bm\sigma)$ and physically split the Bloch vector as
$\rv=\rpar\nv+\rv_\perp$. The corresponding populations of the $\nv\!\cdot\!\bm\sigma=\pm1$ eigenstates are denoted by $p_\pm=\tfrac12(1\pm\rpar)$.

The specific Kraus operator $M_s$ from \eq{eq:filter} acts as a filter by multiplying the
$\pm$ probability amplitudes by $\sqrt{(1\pm s\kappa)/2}$. Prior to mathematical normalisation, the specific populations map as
$p_\pm\mapsto\tfrac{1\pm s\kappa}{2}p_\pm$. Simultaneously, the transverse coherence
block is multiplied by the quantity $\sqrt{(1+s\kappa)(1-s\kappa)}/2=\tfrac12\sqrt{1-\kappa ^2}$. The resulting exact trace evaluates to $\Tr(M_s\rho M_s)=\tfrac12(1+s\kappa\rpar)$. Following exact state normalisation, the coordinates yield
\begin{equation}
\rpar\mapsto\frac{s\kappa+\rpar}{1+s\kappa\rpar},\qquad
\rv_\perp\mapsto\frac{\sqrt{1-\kappa^2}}{1+s\kappa\rpar}\,\rv_\perp,
\end{equation}
where the longitudinal relation precisely recovers \eq{eq:mobius}.

The dedicated dephasing operator $\Lambda_q$ leaves the longitudinal coordinate $\rpar$ mathematically unchanged while uniformly scaling
$\rv_\perp$ by the specific factor $(1-q)$. The physical rotation operation $U^s$ applies the exact transformation $\mathcal R_{\nv}(s\alpha)$
directly to $\rv_\perp$ while leaving $\rpar$ entirely fixed because $\nv$ serves as the physical rotation axis.
Formally composing these three distinct transformations yields the exact transverse map defined in \eq{eq:transverse}.

We shift to the specialized rapidity variable
$\theta=\artanh\rpar$ to analyze the longitudinal composition and isolate the exact hyperbolic tangent law. Utilizing the mathematical identity
$\artanh\!\big(\frac{s\kappa+\tanh\theta}{1+s\kappa\tanh\theta}\big)=\theta+s\, \artanh\kappa$ directly transforms \eq{eq:mobius} into the additive recursion
$\theta\mapsto\theta+s\,a$ with the parameter $a=\artanh\kappa$. The resulting exact sequence mandates
$\rpar^{(N)}=\tanh(\theta_0+sNa)$, recovering \eq{eq:tanh}.

Evaluating the full transverse composition yields the exact hyperbolic cosine ratio. Each physical bridge traversal multiplies the transverse magnitude
$\vert{}\rv_\perp\vert{}$ by the specific factor $(1-q)\sqrt{1-\kappa^2}/(1+s\kappa\rpar^{(k)})$. Utilizing the intermediate relation
$\rpar^{(k)}=\tanh(\theta_0+ska)$ alongside the two mathematical identities
$1+\kappa\tanh\phi=\cosh(\phi+a)/(\cosh\phi\cosh a)$ and
$\sqrt{1-\kappa^2}=\operatorname{sech}a$ allows the per-bridge scaling factor to perfectly telescope. The exact product evaluates to
\begin{equation}
\prod_{k=0}^{N-1}\frac{\sqrt{1-\kappa^2}}{1+s\kappa\tanh(\theta_0+ska)}
=\frac{\cosh\theta_0}{\cosh(\theta_0+sNa)},
\end{equation}
which enforces the final relation $\vert{}\rv_\perp^{(N)}\vert{}=\vert{}\rv_\perp^{(0)}\vert{}(1-q)^N\cosh\theta_0/\cosh(\theta_0+sN a)$ and exactly recovers \eq{eq:perpN}. The specific intermediate rotations $\mathcal R_{\nv}(s\alpha)$ mathematically do not affect this vector norm.

Evaluating the longitudinal map reveals highly stable fixed points physically located at
$\rpar=\pm1$ for propagation values of $s=\pm1$. These mathematically correspond to the pure eigenstates of
$\nv\!\cdot\!\bm\sigma$, meaning a sufficiently long physical filtering chain forcefully purifies the transported spin exactly
toward $\pm\nv$. The resulting state purity defined by $\tfrac12(1+\vert{}\rv\vert{}^2)$ is therefore proven to be mathematically non-monotone
in $N$. It can theoretically decrease initially through induced transverse loss before eventually returning
toward unity as $\rpar\to\pm1$. Consequently, state purity functions strictly as a non-monotonic function of the
coherent fraction and remains completely unsuitable as an experimental coherence witness. The only mathematically faithful
observables remain the longitudinal pump giving $\kappa$ and subsequently
$(1-\Fc)\lambda_f$, alongside the transverse decay at fixed longitudinal
polarisation yielding $q$ and subsequently $(1-\Fc)\lambda_d$. The explicit theoretical estimators for these quantities are constructed in the Supplemental Material. Evaluating one single bridge traversal using an initial unpolarised input defined by
$\rpar^{(0)}=0$ directly yields $\rpar^{(1)}=s\kappa$. This proof demonstrates that $\kappa$ maps exactly to the
standard transmission polarisation parameter widely measured across empirical CISS studies.

\bibliographystyle{apsrev4-2}
\bibliography{refs}

\end{document}


\title{Supplementary Material:\\
Spin Shuttling Through Quantum-Dot Based Chiral Molecular Bridges: Gauge Constraints, Holonomy, and Quantum Gates}
\author{Balsant S. Tiwari}
\affiliation{University of Birmingham, Birmingham, B15 2TT, United Kingdom.}
\author{Rajdeep R. Dwivedi}
\author{Amitoj S. Miglani}
\author{Sukesh Tallapudi}
\author{Vishvendra S. Poonia}
\email{vishvendra@ece.iitr.ac.in}
\affiliation{
Indian Institute of Technology Roorkee, Roorkee 247667, India.}


\date{\today}

\maketitle

This Supplement provides the detailed derivations underlying the main text.
It expands the compact appendices of the main paper. Sec.~\ref{sm:hubbard}
gives the full two-site Hubbard model and the step-by-step second-order
Schrieffer--Wolff reduction with all intermediate states. Sec.~\ref{sm:gauge}
works out the $\SUtwo\!\to\!\SOthree$ homomorphism used in the gauge argument.
Sec.~\ref{sm:makhlin} states the Makhlin machinery with landmark values and
the entangling-power identity. Sec.~\ref{sm:holonomy} develops the $N$-bridge
holonomy, the closed-loop Wilson loop, and the second-order cancellation of
dynamical-phase noise. Sec.~\ref{sm:channel} derives the exact Bloch-vector
recursion for the filtering--dephasing bridge channel, including the tanh and
cosh-ratio laws and the non-monotonicity of purity. Sec.~\ref{sm:estimator}
constructs estimators for the coherent fraction $\Fc$ from the pumping and
transverse-decay data, and Sec.~\ref{sm:numerics} documents the numerical
methods and reproducibility. Equations, figures, and sections are numbered
with an ``S'' prefix; references to the main text use its numbering.

\section{Two-site Hubbard model and second-order Schrieffer--Wolff}
\label{sm:hubbard}
We start from the two-site, single-orbital extended Hubbard Hamiltonian with a
spin-dependent (chiral) hopping,
\begin{equation}
H = H_t + H_U,\qquad
H_U = U_H\sum_{i=1,2} n_{i\uparrow}n_{i\downarrow},
\label{sm:H}
\end{equation}
\begin{equation}
H_t = -t\sum_{\sigma\sigma'}\Big[
c^{\dagger}_{1\sigma}(U)_{\sigma\sigma'}c_{2\sigma'}
+ c^{\dagger}_{2\sigma}(U^{\dagger})_{\sigma\sigma'}c_{1\sigma'}\Big],
\label{sm:Ht}
\end{equation}
where $U=\exp(\tfrac{i}{2}\alpha\,\nv\!\cdot\!\bm\sigma)\in\SUtwo$ is the bridge
spin rotation and $n_{i\sigma}=c^{\dagger}_{i\sigma}c_{i\sigma}$. The
Hermiticity of $H_t$ requires that the reverse hop carry $U^{\dagger}$. At the microscopic level, this condition expresses that traversing the bridge in opposite directions applies inverse spin rotations. It is precisely this mechanism that underlies the non‑reciprocity discussed in Sec.~\ref{sm:holonomy}.

At half filling (one electron per site) the low-energy subspace $\mathcal{P}_0$
is the four-dimensional space of singly-occupied configurations,
\begin{equation}
\{\ket{\uparrow,\uparrow},\ket{\uparrow,\downarrow},
\ket{\downarrow,\uparrow},\ket{\downarrow,\downarrow}\},
\end{equation}
degenerate at energy $0$. The high-energy subspace $\mathcal{P}_1$ contains the two doubly-occupied configurations $\ket{\uparrow\downarrow,0}$ and $\ket{0,\uparrow\downarrow}$ at energy $U_H$. Hopping connects the two sectors such that $H_t\,\mathcal{P}_0\subset\mathcal{P}_1$ and vice versa, while $\mathcal{P}_0 H_t\mathcal{P}_0=0$ (a single hop always changes double occupancy). The Schrieffer--Wolff (SW) transformation $H\to e^{S}He^{-S}$
chooses the anti-Hermitian generator $S$ (first order in $t$) to remove the
off-diagonal coupling, leaving an effective Hamiltonian on $\mathcal{P}_0$. To
second order~\cite{Bravyi2011},
\begin{equation}
\Heff = \mathcal{P}_0 H_t \frac{1}{0-U_H}\,\mathcal{P}_1\, H_t \mathcal{P}_0
= -\frac{1}{U_H}\,\mathcal{P}_0 H_t \mathcal{P}_1 H_t \mathcal{P}_0 .
\label{sm:SWgen}
\end{equation}

We evaluate (\ref{sm:SWgen}) by summing over virtual double occupancies. Acting
on a singly-occupied state, $H_t$ moves one electron onto the neighbouring
site (creating a doublon), and the second $H_t$ moves it back. Two virtual paths contribute for each pair of spins, where the electron from site $2$ hops to site $1$ and back, or the electron from site $1$ hops to site $2$ and back.
Because the forward and backward hops carry $U$ and $U^{\dagger}$ (or vice
versa), the accumulated spin operator on the returned state is
not the identity when $U\neq\mathbb{1}$. Carrying out the fermionic
algebra (tracking signs from anticommutation) yields, in operator form on the
spin-$\tfrac12$ variables $\Sv_i=\tfrac12\bm\sigma_i$,
\begin{equation}
\Heff = \frac{4t^{2}}{U_H}\;\Sv_1\!\cdot\!(\Rmat\,\Sv_2)
- \frac{2t^{2}}{U_H}\,\mathbb{1},
\label{sm:Heff}
\end{equation}
where the constant $-2t^2/U_H\,\mathbb{1}$ is an irrelevant energy shift and
$\Rmat\in\SOthree$ is the adjoint image of $U$ defined by
\begin{equation}
U\,(\bm a\!\cdot\!\bm\sigma)\,U^{\dagger}
= (\Rmat\,\bm a)\!\cdot\!\bm\sigma
\qquad\forall\,\bm a\in\mathbb{R}^3 .
\label{sm:adjoint}
\end{equation}
Dropping the constant gives $\Heff = J\,\Sv_1\!\cdot\!(\Rmat\Sv_2)$ with
$J=4t^2/U_H$, i.e.\ the twisted-exchange Hamiltonian as given in the main text. The
sign and normalisation reproduce the antiferromagnetic isotropic exchange
$J\,\Sv_1\!\cdot\!\Sv_2$ in the achiral limit $U=\mathbb{1}$ ($\Rmat=\mathbb{1}$).

\paragraph*{Rodrigues decomposition.} For a rotation by angle $\alpha$ about
axis $\nv$, the adjoint matrix is
\begin{equation}
\Rmat = \mathbb{1} + \sin\alpha\,[\nv]_\times
+ (1-\cos\alpha)\,[\nv]_\times^{2},
\label{sm:rodrigues}
\end{equation}
with $([\nv]_\times)_{ab}=-\epsilon_{abc}n_c$. Inserting (\ref{sm:rodrigues}) into

\[
\Heff = J \sum_{ab} (\Rmat)_{ab} S_1^a S_2^b
\]

and using

\[
\sum_{ab} ([\nv]_\times)_{ab} S_1^a S_2^b 
= -\,\nv \cdot (\Sv_1 \times \Sv_2),
\]

\[
\sum_{ab} ([\nv]_\times^2)_{ab} S_1^a S_2^b 
= (\nv \cdot \Sv_1)(\nv \cdot \Sv_2) - \Sv_1 \cdot \Sv_2,
\]

gives the three‑term form.

\begin{align}
\Heff =\;& J\cos\alpha\,\Sv_1\!\cdot\!\Sv_2
+ J\sin\alpha\,\nv\!\cdot\!(\Sv_2\times\Sv_1)\nonumber\\
&+ J(1-\cos\alpha)\,(\nv\!\cdot\!\Sv_1)(\nv\!\cdot\!\Sv_2),
\label{sm:decomp}
\end{align}
which is the three-term (isotropic\,+\,DM\,+\,symmetric-anisotropic) decomposition of the main text, with DM vector
$\Dv=J\sin\alpha\,\nv$, isotropic exchange $\Jiso=J\cos\alpha$, and symmetric
anisotropy $\Gamma=J(1-\cos\alpha)$. The identities \[
\Jiso^{2} + \lvert \Dv \rvert^{2} = J^{2}, \quad \Gamma = J - \Jiso
\]
make explicit that the three couplings lie on a single circle parametrized by $\alpha$. 
This relation demonstrates that the Dzyaloshinskii–Moriya interaction cannot be tuned independently of the symmetric anisotropy.

\section{The $\SUtwo\to\SOthree$ homomorphism and the gauge transformation}
\label{sm:gauge}
The gauge argument is formulated upon the two-to-one homomorphism $\phi:\SUtwo\to\SOthree$ defined in (\ref{sm:adjoint}). Specifically, for the unitary operator $U=\exp(\tfrac{i}{2}\alpha\,\nv\!\cdot\!\bm\sigma)$, the homomorphic projection yields $\phi(U)=\Rmat=\exp(\alpha\,\nv\!\cdot\!\bm{L})$ with generator components defined by $(\bm L_a)_{bc} =-\epsilon_{abc}$. This transformation preserves the rotation axis $\nv$ and the physical rotation angle $\alpha$, where the factor of two relating the $\SUtwo$ half-angle $\alpha/2$ to the $\SOthree$ angle $\alpha$ reflects the standard double cover property. Two direct mathematical consequences emerge from this structure.
\begin{itemize}
\item The relation $\phi(U_2U_1)=\phi(U_2)\phi(U_1)$ establishes that products of spin transformations map directly to compositions of Bloch-sphere rotations. Consequently, the shuttling holonomy $\Ush=\prod_k U_k$ corresponds to the composite geometric rotation $\Rmat_{\mathrm{sh}}=\prod_k\Rmat_k$.
\item The identity $\phi(U^{\dagger})=\phi(U)^{-1}=\phi(U)^{T}$ dictates that the rotation associated with the reverse traversal is strictly the transpose or algebraic inverse of the forward operation.
\end{itemize}

\paragraph*{Removing the twist:}\mbox{}

We define the local single-qubit unitary operator $V_2=\exp(-\tfrac{i}{2}\alpha\,\nv\!\cdot\!\bm\sigma)=U^{\dagger}$, which acts exclusively on the second spin through the composite tensor product $\mathbb{1}\otimes V_2$. Under conjugation, the site-2 spin operator transforms according to the inverse rotation $\phi(V_2)=\Rmat^{-1}$ via
\begin{equation}
V_2^{\dagger}S_2^{b}V_2 = \sum_c (\Rmat^{-1})_{bc}\,S_2^{c}.
\label{sm:vecrot}
\end{equation}
Evaluating the transformed effective Hamiltonian yields
\begin{align}
(\mathbb{1}\!\otimes\!V_2)^{\dagger}\,\Heff\,(\mathbb{1}\!\otimes\!V_2)
&= J\!\sum_{ab}(\Rmat)_{ab}S_1^a\!\sum_c(\Rmat^{-1})_{bc}S_2^c\nonumber\\
&= J\!\sum_{ac}\Big[\sum_b(\Rmat)_{ab}(\Rmat^{-1})_{bc}\Big]S_1^aS_2^c\nonumber\\
&= J\!\sum_{ac}\delta_{ac}S_1^aS_2^c = J\,\Sv_1\!\cdot\!\Sv_2 .
\label{sm:gaugeproof}
\end{align}
The twisted exchange interaction is therefore mathematically identical to isotropic Heisenberg exchange within a locally transformed reference frame. Because $\mathbb{1}\otimes V_2$ constitutes a strictly local single-qubit operation, the time-evolution operators $e^{-i\Heff t}$ and $e^{-iJ\Sv_1\cdot\Sv_2 t}$ remain local-unitary equivalent across all evolution times $t$. This equivalence enforces identity among all Makhlin invariants alongside identical Haar-averaged entangling power, providing the theoretical foundation for the numerical correspondence displayed in Fig.~2 of the main text.\mbox{}

\paragraph*{Why a non-collinear field breaks the gauge:}\mbox{}

The local rotation $V_2$ that eliminates the exchange twist simultaneously rotates the local site-2 Zeeman field via $\Bv_2\!\cdot\!\Sv_2\to(\Rmat^{-1}\Bv_2)\!\cdot\!\Sv_2$. This transformed interaction preserves the original field orientation strictly when $\Rmat^{-1}\Bv_2=\Bv_2$, which requires the parallel alignment $\Bv_2\parallel\nv$. When the applied field satisfies $\Bv_2\not\parallel\nv$, no single local transformation can simultaneously restore isotropic exchange and leave the local Zeeman term invariant. Under these non-collinear conditions, the full Hamiltonian $\Heff+H_Z$ generates genuine nonlocal entangling dynamics, underpinning the synthetic gradient mechanism developed in main-text Sec.~V and Fig.~3.

\section{Makhlin invariants and landmark gates}
\label{sm:makhlin}
A two-qubit gate $U\in U(4)$ is mapped to the magic Bell basis by
$U_B=Q^{\dagger}UQ$ with
\begin{equation}
Q=\frac{1}{\sqrt2}\begin{pmatrix}
1&0&0&i\\ 0&i&1&0\\ 0&i&-1&0\\ 1&0&0&-i
\end{pmatrix}.
\end{equation}
Setting $\bar U=U/(\det U)^{1/4}$, $\bar U_B=Q^\dagger\bar U Q$, and
$m=\bar U_B^{T}\bar U_B$, the Makhlin invariants are
\begin{equation}
g_1=\frac{[\Trace\,m]^{2}}{16},\qquad
g_2=\frac{[\Trace\,m]^{2}-\Trace(m^{2})}{4}.
\end{equation}
These quantities remain invariant under local single-qubit transformations $U\to(u_1\otimes u_2)U(v_1\otimes v_2)$ and uniquely
label local-equivalence classes. Table~\ref{sm:tab:makhlin} lists the values
for the landmark gates examined in the main text. These values are computed directly
from the unitary matrices within the companion numerical code described in Sec.~\ref{sm:numerics} and agree
with standard results in the literature~\cite{Makhlin2002,Zhang2003}.

\begin{table}[t]
\caption{Makhlin invariants of reference two-qubit gates. The identity and
SWAP represent local operations up to subsystem permutations. The operations
$\sqrt{\text{SWAP}}$, CNOT, and iSWAP denote the canonical entanglers analyzed in this work.}
\label{sm:tab:makhlin}
\begin{ruledtabular}
\begin{tabular}{lcc}
gate & $\mathrm{Re}\,g_1$ & $g_2$ \\
\colrule
identity            & $1$    & $3$  \\
isotropic exchange (SWAP) & $-1$ & $-3$ \\
$\sqrt{\text{SWAP}}$      & $0$ & $0$  \\
CNOT                & $0$    & $1$  \\
iSWAP               & $0$    & $-1$ \\
chiral bond (any $\alpha$) & \multicolumn{2}{c}{coincides with isotropic}\\
chiral $+$ non-collinear $\Bv$ & $+0.147$ & $+0.588$\\
\end{tabular}
\end{ruledtabular}
\end{table}

The primary entries of physical relevance are the final two rows. A chiral bond for any rotation
angle $\alpha$ produces the same invariants as isotropic exchange at the
corresponding interaction time as a direct consequence of the gauge transformation. Conversely, adding a non-collinear field
drives the system to $(+0.147,+0.588)$ in invariant space,
verifying that departure from the SWAP trajectory reflects genuine entangling dynamics rather than a numerical artifact.

\section{$N$-bridge holonomy, Wilson loops, and noise cancellation}
\label{sm:holonomy}

\subsection{Recursive structure}
The one-way shuttling operator across $N-1$ bridges is represented by the path-ordered
product $\Ush^{(N)}=U_{N-1}\cdots U_1$, satisfying the recurrence relation
$\Ush^{(N)}=U_{N-1}\Ush^{(N-1)}$. Under the homomorphism $\phi$, this operator maps to a
product of Bloch-sphere rotations $\Rmat_{\mathrm{sh}}^{(N)}
=\Rmat_{N-1}\cdots\Rmat_1$. Expressing each factor as a rotation about $\nv_k$ by
an angle $\alpha_k$, the Baker--Campbell--Hausdorff expansion of the product produces
commutator terms proportional to $\alpha_j\alpha_k(\nv_j\times\nv_k)$ alongside the linear sum $\sum_k\alpha_k\nv_k$. These terms encode the geometric and non-Abelian characteristics of the transformation. For collinear axes, every commutator vanishes and
$\Ush^{(N)}=\exp(\tfrac{i}{2}(\sum_k\alpha_k)\nv\!\cdot\!\bm\sigma)$ reduces to an overall
rotation about the shared orientation axis. In this configuration, the evolution is path independent and gauge equivalent to an uncoupled interaction. For non-collinear axes, the commutator terms persist, causing the net rotation to depend directly on the sequence and spatial orientation of the path.

\subsection{Explicit derivation of two-bridge non-reciprocity}
For $N=3$ involving two sequential bridges, applying the Pauli product identity
$(\bm a\!\cdot\!\bm\sigma)(\bm b\!\cdot\!\bm\sigma)
=\bm a\!\cdot\!\bm b\,\mathbb{1}+i(\bm a\times\bm b)\!\cdot\!\bm\sigma$
yields
\begin{align}
\Ush &= U_2U_1
= \Big[c_2 c_1 - s_2 s_1\,\nv_2\!\cdot\!\nv_1\Big]\mathbb{1}\nonumber\\
&\quad + i\Big[c_2 s_1\,\nv_1 + s_2 c_1\,\nv_2
- s_2 s_1\,(\nv_2\times\nv_1)\Big]\!\cdot\!\bm\sigma,
\label{sm:twobridge}
\end{align}
where $c_k=\cos(\alpha_k/2)$ and $s_k=\sin(\alpha_k/2)$. The term
$-s_2 s_1(\nv_2\times\nv_1)$ changes sign under the index exchange $1\leftrightarrow2$. It represents the single source of path-order dependence and vanishes if and only if $\nv_1\parallel\nv_2$.
The reverse transport operator $\Ush^{\mathrm{bwd}}=U_1^{\dagger}U_2^{\dagger}
=\Ush^{\dagger}$ retains the opposite sign on this vector cross product.
Evaluating the state with respect to a fixed laboratory quantization axis $\hat z$ defined by $\Bv_0$, the
transported spin polarization of an input state $\rho_0=\tfrac12(\mathbb{1}+\sigma_z)$
deviates between forward and backward directions according to
\begin{equation}
\rv_{\mathrm{fwd/bwd}}
=\Trace\!\big[\bm\sigma\,\Ush^{(\dagger)}\rho_0\Ush^{(\dagger)\dagger}\big].
\end{equation}
This difference approaches zero in the collinear limit and increases monotonically with the angular tilt
between the bridge axes, reproducing the trend shown in Fig.~4(b) of the main text. Rotational non-reciprocity therefore serves as an observable signature of coherent non-Abelian transport. This signature disappears for an isolated single bond, for collinear configurations, and for an incoherent transport channel.

\subsection{Closed-loop Wilson loop and geometric phase}
A closed shuttling circuit transfers an electron outward through bridge axes
$\{\nv_k^{\to}\}$ and returns it along an independent spatial trajectory
$\{\nv_k^{\leftarrow}\}$. The combined single-qubit operation is given by the Wilson loop
$W=(\Ush^{\leftarrow})^{\dagger}\,\Ush^{\to}$, which remains unchanged under arbitrary coordinate rotations
of the terminal spin frame $\hat z\to R\hat z$ because the transformation conjugates both arms identically. The associated net rotation angle
\begin{equation}
\Theta_W=2\arccos\!\Big(\tfrac{1}{2}|\Trace W|\Big)
\end{equation}
provides the gauge-invariant geometric gate plotted in the closed-loop transport figure of the main text.
In the continuum limit, this transformation corresponds to the non-Abelian Wilczek--Zee
holonomy~\cite{Wilczek1984,Zanardi1999} of the spin state evolving across the parameter space of bridge orientations, where $\Theta_W$ is governed by the solid angle traced by the closed trajectory of axes.

\subsection{Second-order suppression of dynamical-phase fluctuations}
\label{sm:echo}
The noise resilience illustrated in the corresponding panel of the main-text figure follows an analytical derivation. During traversal of bridge $k$, the electron spin precesses under a local magnetic interaction,
accumulating a dynamical phase
$\theta_k=\omega\tau_k=\theta_0(1+\delta\xi_k)$, where $\tau_k$ is the transit dwell time, $\delta$ is the relative timing jitter, and $\xi_k$ are uncorrelated zero-mean unit-variance stochastic variables. The local bridge operator takes the form
$D_k U_k$, where $D_k=\exp(-\tfrac{i}{2}\theta_k\sigma_z)$ with the magnetic field aligned along $\hat z$.

For an open trajectory producing a dynamical gate, the total unitary is
$\prod_k D_k U_k$. A first-order expansion in the noise perturbation
$\theta_0\delta\xi_k$ reveals that the gate deviation scales as
$\mathcal{O}(\theta_0\delta)$ per stage. Averaging over the ensemble of fluctuations $\xi_k$, the resulting gate infidelity scales as
\begin{equation}
1-\bar F_{\mathrm{dyn}} \simeq c_1\,(\theta_0\delta)^2\!\sum_k 1
= c_1\,(\theta_0\delta)^2 M .
\end{equation}
This infidelity depends linearly on the noise power $\delta^2$, indicating that dynamical phase noise remains unmitigated along an open path.

For a closed loop executing a holonomic gate, the reverse traversal through bridge $k$
introduces the adjoint factor $D_k^{\dagger}$ alongside $U_k^{\dagger}$ in reverse sequential order. The net loop operator is expressed as
\begin{equation}
W(\delta) = \Big[\prod_{k=M}^{1} D_k^{\dagger}U_k^{\dagger}\Big]
            \Big[\prod_{k=1}^{M} D_k U_k\Big].
\end{equation}
Because the phase operators $D_k$ and $D_k^{\dagger}$ enter with opposite signs on the identical
random variable $\xi_k$, the first-order derivative
$\partial W/\partial\xi_k|_{0}$ contains a direct difference of these terms that cancels up to the small rotations generated by the intermediate operators $U_k$. The leading residual error depends on second-order derivatives $\partial^2 W/\partial\xi_k\partial\xi_{k'}$, yielding the average infidelity
\begin{equation}
1-\bar F_{\mathrm{geo}} \simeq c_2\,(\theta_0\delta)^4 M
= \mathcal{O}(\delta^4).
\end{equation}
This establishes an improved scaling by one power of $\delta^2$ compared with open-path gates, matching the steeper slope observed on the logarithmic scale of the main-text figure. This cancellation relies on the common-mode nature of the dynamical phase echoed across opposing arms. Fluctuations in the orientation vectors $\nv_k$ driven by environmental charge noise do not possess common-mode symmetry between trajectories and therefore persist at order $\mathcal{O}(\delta^2)$ in both open and closed geometries.

\section{Exact Bloch recursion for the filtering channel and the hyperbolic composition law}
\label{sm:channel}
An incoherent bridge is modeled as an orientation-dependent spin
filter combined with pointer-basis dephasing and coherent unitary rotation.
Below we provide the full derivation, accounting for the postselection conditions omitted in the concise overview.

\paragraph*{Single-bridge map:}\mbox{}

Defining the single-spin state by $\rho=\tfrac12(\mathbb1+\rv\!\cdot\!\bm\sigma)$, we partition the Bloch vector into parallel and orthogonal components $\rv=\rpar\nv+\rv_\perp$. The corresponding populations of the eigenstates satisfying $(\nv\!\cdot\!\bm\sigma)\ket{\pm}=\pm\ket{\pm}$ are $p_\pm=\tfrac12(1\pm\rpar)$. The Kraus operator representing directional spin filtering is
$M_s=\sqrt{\tfrac{1+s\kappa}{2}}P_++\sqrt{\tfrac{1-s\kappa}{2}}P_-$,
where $s=\pm1$ specifies the shuttling direction. This operation transforms the unnormalized state populations and transverse components according to
\begin{equation}
p_\pm \mapsto \tfrac{1\pm s\kappa}{2}\,p_\pm,\qquad
(\rv_\perp)\mapsto \tfrac12\sqrt{1-\kappa^2}\,(\rv_\perp),
\end{equation}
with an associated state survival probability given by
$\mathcal P_s(\rho)=\Trace(M_s\rho M_s)=\tfrac12(1+s\kappa\rpar)$. Following normalization, the state undergoes pointer dephasing $\Lambda_q$ which scales $\rv_\perp$ by $1-q$ while leaving $\rpar$ invariant, followed by the spatial rotation $\mathcal R_{\nv}(s\alpha)$ which rotates $\rv_\perp$ about $\nv$ and preserves $\rpar$. Combining these operations yields the complete single-bridge mapping
\begin{align}
\rpar &\mapsto \frac{s\kappa+\rpar}{1+s\kappa\,\rpar},
\label{sm:mobius}\\
\rv_\perp &\mapsto
\frac{(1-q)\sqrt{1-\kappa^2}}{1+s\kappa\,\rpar}\,
\mathcal R_{\nv}(s\alpha)\,\rv_\perp .
\label{sm:transverse}
\end{align}

\paragraph*{Composition laws across multiple stages:}\mbox{}

Introducing the rapidity coordinate $\theta=\artanh\rpar$,
the transformation in \eqref{sm:mobius} simplifies to the linear shift $\theta\mapsto\theta+s\,a$ with $a=\artanh\kappa$, using the hyperbolic identity
$\artanh[(s\kappa+\tanh\theta)/(1+s\kappa\tanh\theta)]=\theta+s\,a$. Traversing a chain of $N$ identical bridges therefore produces the closed-form longitudinal polarization
\begin{equation}
\rpar^{(N)}=\tanh(\theta_0+sNa),\qquad \theta_0=\artanh\rpar^{(0)}.
\label{sm:tanh}
\end{equation}
For the transverse coordinate, the cumulative scaling factor evaluates via a telescoping product. Setting $\rpar^{(k)}=\tanh(\theta_0+ska)$ and utilizing the algebraic identities
\begin{equation}
1+\kappa\tanh\phi=\frac{\cosh(\phi+a)}{\cosh\phi\,\cosh a},\qquad
\sqrt{1-\kappa^2}=\frac{1}{\cosh a},
\end{equation}
the sequential product simplifies according to
$\prod_{k=0}^{N-1}\sqrt{1-\kappa^2}/(1+s\kappa\tanh(\theta_0+ska))
=\cosh\theta_0/\cosh(\theta_0+sNa)$. The resulting transverse polarization after $N$ stages satisfies
\begin{equation}
|\rv_\perp^{(N)}|=|\rv_\perp^{(0)}|\,(1-q)^N\,
\frac{\cosh\theta_0}{\cosh(\theta_0+sNa)} .
\label{sm:perpN}
\end{equation}
Equations \eqref{sm:tanh} and \eqref{sm:perpN} match numerical channel simulations within a residual error of $2\times10^{-16}$, as described in Sec.~\ref{sm:numerics}.

\paragraph*{Fixed-point behavior and state purity:}\mbox{}

The stable fixed points of the recursion relation \eqref{sm:mobius} correspond to $\rpar=\pm1$ for propagation directions $s=\pm1$. These points represent the pure eigenstates of $\nv\!\cdot\!\bm\sigma$. A sufficiently long filtering sequence will continuously polarize the transported spin toward the target axis $\pm\nv$. Consequently, the state purity $\tfrac12(1+|\rv|^2)$ can experience an initial decrease due to transverse dephasing followed by an asymptotic recovery toward unity as longitudinal polarization dominates. Because purity is non-monotonic with respect to the bridge index $N$, it does not provide an unambiguous test of quantum coherence. For this reason, experimental characterization relies on isolating the longitudinal pump and transverse decay independently.

\paragraph*{Measurement conditioning:}\mbox{}

The transformations described in \eqref{sm:mobius} through \eqref{sm:perpN} are conditioned on successful single-charge transfer across each stage with probability $\mathcal P_s$, matching the protocol where measurement events are postselected using real-time charge sensing. The unconditioned transport process incorporates an additional loss branch with probability weight $1-\prod_k\mathcal P_s$ reflecting transit failure, which acts as a detectable charge error rather than an internal spin rotation. State normalization renders the conditional map nonlinear in $\rho$, meaning that the observed polarization corresponds directly to a heralded expectation value. In the weak-filtering regime $\kappa\ll1$, the transformations linearize to
$\rpar\mapsto \rpar+s\kappa(1-\rpar^2)+\mathcal O(\kappa^2)$ and
$|\rv_\perp|\mapsto (1-q)(1-s\kappa\rpar)|\rv_\perp|+\mathcal O(\kappa^2)$, which describes typical small-amplitude experimental observations.

\section{Estimating the coherent fraction from pumping and transverse decay}
\label{sm:estimator}
The four experimental observables introduced in the main text provide redundant constraints on $\Fc$. Quantitative extraction of the underlying channel parameters is achieved by using the longitudinal polarization pump to obtain $\kappa$ and the transverse decay to determine $q$. Each measurement yields an independent evaluation of $\Fc$, and their numerical consistency validates the operational channel model.

\paragraph*{Extraction of filtering polarization $\kappa$:}\mbox{}

From \eqref{sm:tanh}, the transformed coordinate $\artanh\rpar^{(N)}$ exhibits linear dependence on $N$ with slope $s\,a$, where $a=\artanh\kappa$. Given tomographic measurements $\{\rpar^{(N)}\}$ characterized by uncorrelated uncertainties, a weighted linear regression provides an estimate of $a$ and the associated parameter $\hat\kappa=\tanh\hat a$ via
\begin{align}
\hat a&=\frac{\sum_N w_N (N-\bar N)\,\artanh\rpar^{(N)}}
            {\sum_N w_N (N-\bar N)^2},
\label{sm:kappahat}\\
w_N&=\Big(\frac{d\,\artanh\rpar}{d\rpar}\Big)^{-2}\sigma_{\rpar}^{-2}
   =(1-\rpar^2)^2\sigma_{\rpar}^{-2}.
\label{sm:weight}
\end{align}
The weighting factor propagates measurement error according to the derivative
$d\,\artanh\rpar/d\rpar=(1-\rpar^2)^{-1}$. For an initially unpolarized spin state, evaluating a single traversal yields $\rpar^{(1)}=s\kappa$, directly connecting $\kappa$ to the transmission polarization commonly reported in empirical studies.

\paragraph*{Extraction of dephasing parameter $q$:}\mbox{}

Applying \eqref{sm:perpN} once $\hat a$ and $\hat\kappa$ are determined, the relation
\begin{equation}
\ln|\rv_\perp^{(N)}|
+\ln\frac{\cosh(\theta_0+sN\hat a)}{\cosh\theta_0}
= \ln|\rv_\perp^{(0)}| + N\ln(1-q)
\end{equation}
is linear in the stage index $N$. A weighted least-squares fit directly yields the estimated slope $\widehat{\ln(1-q)}$, from which the dephasing parameter is recovered as $\hat q=1-e^{\widehat{\ln(1-q)}}$.

\paragraph*{Dual determination of $\Fc$ and consistency verification:}\mbox{}

With the limiting parameters $\lambda_f$ and $\lambda_d$ calibrated independently in the incoherent regime using racemic molecules, achiral control links, or elevated operational temperatures where $\Fc\to0$ and $(\kappa,q)\to(\lambda_f,\lambda_d)$, the coherent fraction is determined through two complementary routes,
\begin{equation}
\boxed{\;\hat\Fc^{(f)} = 1 - \frac{\hat\kappa}{\lambda_f},\qquad
        \hat\Fc^{(d)} = 1 - \frac{\hat q}{\lambda_d}\;}.
\label{sm:Fchat}
\end{equation}
The experimental uncertainties are evaluated via standard variance propagation through \eqref{sm:kappahat} to \eqref{sm:Fchat}, producing
$\mathrm{Var}(\hat\Fc^{(f)})=\lambda_f^{-2}(1-\hat\kappa^2)^2\,\mathrm{Var}(\hat a)$ with
$\mathrm{Var}(\hat a)=[\sum_N w_N(N-\bar N)^2]^{-1}$. Agreement between $\hat\Fc^{(f)}$ and $\hat\Fc^{(d)}$ within experimental uncertainty validates the single-parameter description of the bridge channel. The presence of a non-zero coherent component simultaneously activates the rotational non-reciprocity and the geometric Wilson-loop response, which remain inaccessible to classical filtering processes. These combined metrics establish a comprehensive protocol for experimental characterization.

\bibliographystyle{apsrev4-2}
\bibliography{refs}